\documentclass[12pt]{article}

\usepackage{amsfonts}
\usepackage{pstricks}
\usepackage{pst-node}
\usepackage{epsfig}
\usepackage{amsmath,amssymb,latexsym}

\usepackage{import}

\usepackage{enumerate}

\usepackage{physics}

\usepackage[lmargin=1.0in, rmargin=1.0in, tmargin=1.0in, bmargin=1.0in]{geometry}

\usepackage{cancel}

\usepackage{empheq}

\usepackage{comment}

\usepackage{setspace}
\begin{document}

%--------------------------------------------
% several abbreviations
\def\ba{\begin{eqnarray}}
\def\ea{\end{eqnarray}}
\def\w{\wedge}
%--------------------------------------------

%\setstretch{1.0}

%\begin{titlepage}
%\title{\bf Metric formulation of teleparallel geometries}
\title{\bf General teleparallel metrical geometries}
\author{Muzaffer Adak$^{1,2,}$\footnote{madak@pau.edu.tr},  Tekin Dereli$^{3,4,}$\footnote{tdereli@ku.edu.tr}, Tomi S. Koivisto$^{5,6,}$\footnote{tomi.koivisto@ut.ee} , Caglar Pala$^{1,5,7,}$\footnote{caglar.pala@gmail.com}\\
 {\small $^1$ Department of Physics, Faculty of Science, Pamukkale University, 20017 Denizli, Türkiye} \\
 {\small $^2$ Department of Physics, Istanbul Technical University, Maslak 34469 Istanbul, Türkiye} \\
{\small $^3$ Department of Basic Sciences, Faculty of Engineering and Natural Sciences, Maltepe University,} \\
{\small 34857 Maltepe, İstanbul, Türkiye} \\
 {\small $^4$ Department of Physics, Koç University, 34450 Sarıyer, Istanbul,Türkiye} \\
  {\small $^5$ Laboratory of Theoretical Physics, Institute of Physics, University of Tartu, Tartu, Estonia} \\
  {\small $^6$ National Institute of Chemical Physics and Biophysics, R\"avala pst. 10, 10143 Tallinn, Estonia} \\
   {\small $^7$ Department of Physics, Faculty of Science, Erciyes University, 38280 Kayseri, Türkiye} 
 }
 
  \vskip 1cm
\date{\today}
\maketitle

 \thispagestyle{empty}

\begin{abstract}
 \noindent

In the conventional formulation of general relativity, gravity is represented by the metric curvature of Riemannian geometry. There are also alternative formulations in flat affine geometries, wherein the gravitational dynamics is instead described by torsion and nonmetricity. These so called general teleparallel geometries may also have applications in material physics, such as the study of crystal defects. In this work, we explore the general teleparallel geometry in the language of differential forms. We discuss the special cases of metric and symmetric teleparallelisms, clarify the relations between formulations with different gauge fixings and without gauge fixing, and develop a method of recasting Riemannian into teleparallel geometries. As illustrations of the method, exact solutions are presented for the generic quadratic theory in 2, 3 and 4 dimensions.
 
%It is well known that gravity is represented by curvature which is derived analytically from metric in general relativity formulated in the Riemannian geometry. 
%One of alternative gravity models is general teleparallel gravity in which gravity is represented by both torsion and nonmetricity. As a subcase of that, it is represented only by torsion which is called metric (or Weitzenböck) teleparallel gravity. As another subcase, there is also symmetric teleparallel gravity in which gravity is represented only by nonmetricity. We argue that although it can not derived analytically, all three teleparallel geometries can be formulated by one metric tensor like in Riemannian geometry. Mathematical methods and techniques developed here may be used in non-gravitational researches such as crystal defects of material science.  \\

%\vskip 1.0cm
%\bigskip


% 11.15.Kc    (Gauge field theory) Classical and semiclassical techniques
% 04.50.Kd    Modified theories of gravity
% 02.40.Yy    Geometric mechanics
\end{abstract}
%\end{titlepage}

%\setcounter{page}{1}

\section{Introduction}

Einstein's theory of general relativity is mathematically consistent and successful in explaining many observational data. In this theory, gravity is geometrized by associating it with the curvature calculated from the metric of Riemannian spacetime. On the other hand, there are strong motivations to investigate alternative gravity theories, such as the inadequacy of general relativity to explain dark matter and dark energy, the lack of its consistent quantization etc. There are several routes one may take to modify gravity. Interesting possibilities arise in generalised geometries found on routes that lead beyond the Riemannian spacetime. Once we allow an affine connection independently from the Levi-Civita (or Christoffel) connection which is written in terms of metric, we pass to the non-Riemannian geometry defined the triple $(M,g,\nabla)$ where $M$ is a manifold, $g$ is a metric and $\nabla$ is an affine connection. Thus, there are three quantities that we can play with independently; nonmetricity, torsion, curvature, $(Q_{ab}, T^a, R^a{}_b)$, respectively\footnote{Since we use exterior algebra language, in fact we should say nonmetricity 1-form, torsion 2-form, curvature 2-form.} \cite{benn1982,dereli-onder-1996,hehl1995}. 

In this article, we consider teleparallel spacetimes defined by zero curvature. The three cases of interest are: i. general teleparallelism $(Q_{ab} \neq 0, T^a \neq 0, R^a{}_b=0)$ \cite{koivisto-jimenez-2020,Heisenberg:2022mbo}, ii. metric (Weitzenböck) teleparallelism $(Q_{ab} = 0, T^a \neq 0, R^a{}_b=0)$ \cite{yakov-itin-1999,Hohmann:2022mlc}, iii. symmetric teleparallelism $(Q_{ab} \neq 0, T^a = 0, R^a{}_b=0)$ \cite{adak2006,adak2013ijmpa}. There is an extensive literature on the cosmological applications of metric teleparallel modified gravity models \cite{Bahamonde:2021gfp}, and some investigations of symmetric teleparallel modified gravity models have been carried out, but the most general modifications are almost unexplored \cite{Heisenberg:2022mbo}. There is also motivation from the more foundational perspective, 
since
 the rigorous implementation of the principle of relativity, which is lacking in Einstein's original formulation of his theory, requires the extension of the symmetry group by the general linear group, and thereby implies the geometric framework of general teleparallelism \cite{Koivisto:2022nar}. In this article, we begin to systematically explore general teleparallel theories using the language of differential forms.

We will show how given a metric, one can set up examples from each of the three cases of teleparallel geometries by exploiting symmetries. Thus, our point of departure is upon a familiar ground, which we transform to something different without invoking new degrees of freedom in the process\footnote{Also, there are doubts whether teleparallel gravity models without metric equivalents exist, due to foundational problems with extra degrees of freedom \cite{Golovnev:2020zpv,BeltranJimenez:2020fvy}.}.
It will be shown that geometries of the types i. ii. and iii. above can be constructed solely from a metric\footnote{Of course, we do not mean that arbitrary teleparallel geometries could be reduced to metric ones. In an appendix we show that it is (in a certain sense) possible in the type iii. teleparallelism.}, albeit in a non-analytical way unlike the Riemannian geometry which allows the analytical formulation purely in terms of a metric. There have been works on such metric formulations of metric teleparallel and symmetric teleparallel geometries in the literature \cite{yakov-itin-1999,adak2013ijmpa}, but the metric formulation of general teleparallelism is discussed here for the first time. 
Also, we construct an even-parity quadratic Lagrangian in each of the three cases and perform independent variations with respect to the orthonormal coframe, $e^a$, the affine connection 1-form, $\omega^a{}_b$, and the concerned Lagrange multipliers to obtain field equations explicitly in the language of exterior algebra. Since there are examples of metric formulation of Weitzenböck and symmetric teleparallel theories of gravity in the literature we give some concrete examples in two, three and four dimensions for the general teleparallel theory of gravity to make our arguments clear and concrete.

%After discussing the metric formulation of each teleparallel geometry we write down 

Besides gravity, there are other motivations to explore general teleparallel geometries. In studies of photonic crystals, a lattice pattern is basically formed by using two different optical materials. If they are placed periodically in one (or two) space dimension(s), that structure is called one (or two)-dimensional photonic crystal. Of course, there may be three-dimensional photonic crystals. In practice two-dimensional ones which are constructed by arranging very tiny (in nanometer scale) dielectric roads at lattice points are the most commonly investigated. The aim of these searches is mainly to control the behaviour of electromagnetic wave by creating crystal defects in lattice pattern. Defects can be produced by various methods such as by changing the radius or the dielectric constant, by removing a selected rod, by adding a tiny auxiliary extra rod to some main roads etc \cite{kazuaki-sakoda-2005}. On the other hand, in the literature there are works on non-Riemannian formulations of crystal defects relating torsion, curvature and nonmetricity with densities of dislocations, disclinations and metric anomalies, respectively,  \cite{dereli-vercin-1987}-\cite{roychowdhury-gupta-2017}. Thus, mathematical insights and techniques developed here could be applicable in material physics such as the study of crystal defects. For the literature on the non-Riemannian formulation of crystal defects one can consult for the nice paper \cite{roychowdhury-gupta-2017} and the references therein.          

In the following section we summarize our mathematical notations and definitions such as the coordinate frame, orthonormal frame, mixed frame, the decomposition of full (affine) connection, and the variation of a Lagrangian in detail. In subsections of Section \ref{sec:metric-formulation-teleparallel-geometries}, after recalling very briefly the Riemannian spacetime and general theory of relativity, we discuss metric (Weitzenböck), symmetric and general teleparallel spacetimes and vary the concerned Lagrangian for each. We present some classes of exact solutions in two, three and four dimensions for the general teleparallel gravity. We adhere to the exterior algebra of differential forms throughout the paper. Especially when searching for solutions we use the computer algebra system REDUCE \cite{hearn-2004} and its exterior algebra package EXCALC \cite{schrufer-2004}.  In Section \ref{sec:conclusion} we conclude with some discussion.

\section{The mathematical preliminaries} \label{sec:mathematics}

We denote a differentiable metric affine geometry by the triple $\{ M,g,\nabla\}$ where $M$ is the $n$-dimensional orientable and differentiable manifold, $g$ is the (0,2)-type symmetric and non-degenerate metric tensor, $\nabla$ is the affine connection representing the parallel transport of tensors (and also spinors). Let $x^\alpha(p), \ \alpha =\hat{0},\hat{1}, \cdots , \hat{n}-\hat{1}$, be the coordinate functions of the coordinate system at any point $p \in M$. This coordinate system forms the natural reference frame denoted by $ \frac{\partial}{\partial x^\alpha}(p) \equiv \partial_\alpha (p) $, the so-called {\it coordinate frame}. This frame is a set of basis vectors at point $p$ for the tangent space $T_p(M)$. The union of all tangent spaces on $M$ is called the coordinate tangent bundle; $CT(M) = \bigcup_{p \in M}T_p(M)$. Similarly, the differential of coordinate functions $d x^\alpha (p) \equiv e^\alpha(p)$ forms the {\it coordinate (or holonomic) coframe}  at the point $p$ for the cotangent space $T_p^*(M)$. Again the union of all $ T_p^*(M)$ establishes the coordinate cotangent bundle; $CT^*(M) = \bigcup_{p \in M}T_p^*(M)$. Duality between them is formulated by the relation
 \begin{equation}\label{duality1}
          d x^\alpha \left( \partial_\beta \right) = \delta^\alpha_\beta 
 \end{equation}
where $\delta^\alpha_\beta$ is the Kronecker symbol. In the coordinate frame the metric can be expressed in terms of its components by
 \begin{equation}\label{metric1}
   g=g_{\alpha \beta}(x) dx^\alpha \otimes dx^\beta
 \end{equation}
where $\otimes$ denotes the symmetric tensor product, $g_{\alpha \beta}(x) = g_{\beta \alpha}(x)$. We notice that the components of the metric tensor depend on the coordinates, $g(\partial_\alpha , \partial_\beta)=g_{\alpha \beta}(x)$ or $dg_{\alpha \beta} \neq 0$ where $d$ denotes the exterior derivative.

On the other hand, one can always construct an orthonormal frame $X_a,\ a=0,1,\cdots , n-1$, with help of metric. Thus, $X_a$ is dubbed the {\it $g$-orthonormal frame} in which the metric components become $g(X_a,X_b)=\eta_{ab}$ where $\eta_{ab}=\mbox{diag}(-1,1, \cdots , 1)$ is the Minkowski metric. We generally call it as the orthonormal frame in short. The orthonormal frame and coordinate frame are related through n-bein $h^\alpha{}_a$ or its inverse $h^a{}_\alpha$, or vice versa,
  \begin{equation}\label{doublet1}
     X_a (x) = h^\alpha{}_a (x) \partial_\alpha \qquad \Leftrightarrow \qquad \partial_\alpha = h^a{}_\alpha (x) X_a (x)
  \end{equation}
such that $h^\alpha{}_a(x) h^a{}_\beta(x) = \delta^\alpha_\beta$ and $h^a{}_\alpha (x) h^\alpha{}_b (x) = \delta^a_b$. They are elements of the general linear group, $GL(n,\mathbb{R})$. Thus, we can obtain the {\it orthonormal coframe} $e^a$ through the duality relation
 \begin{align}
               e^a(X_b) = \delta^a_b .\label{duality2}
 \end{align}
This is another manifestation of the duality relation (\ref{duality1}). We can always pass from the orthonormal coframe to the coordinate coframe or vice versa by help of the n-bein defined in (\ref{doublet1}) as
  \begin{equation}\label{doublet2}
     dx^\alpha = h^\alpha{}_a(x) e^a(x) \qquad \Leftrightarrow \qquad e^a(x) = h^a{}_\alpha(x) dx^\alpha .
  \end{equation}
While $X_a(p)$ is the orthonormal basis of the tangent space $T_p(M)$, its dual $e^a(p)$ is the orthonormal basis of the cotangent space $T_p^*(M)$ at $p$ of $M$. Consequently, as the union of all $T_p(M)$ with $X_a(p)$ establishes the orthonormal tangent bundle, $OT(M)$, the union of all $T_p^*(M)$ with $e^a(p)$ forms the orthonormal cotangent bundle, $OT^*(M)$. In the orthonormal frame the metric (\ref{metric1}) takes the form
 \begin{equation}\label{metric2}
   g = \eta_{ab} e^a(x) \otimes e^b(x) .
 \end{equation}
Here we pay special attention that the metric components $g(X_a,X_b)=\eta_{ab}$ are independent of the coordinates, that is, $d\eta_{ab}=0$.

In this work we use the language of the exterior algebra in which the coframe is called as the basis 1-form. Accordingly, we call $dx^\alpha$ as the coordinate 1-form and $e^a$ as the orthonormal 1-form (more precisely, metric-orthonormal 1-form). The symbol $d$ is the exterior derivative converting a $p$-form to a $(p+1)$-form. Besides, the exterior derivative of coframe is defined as the anholonomity 2-form. Therefore, since $de^\alpha \equiv d(dx^\alpha) = 0$ because of the Poincar{\'e} lemma, $d^2=0$, $e^\alpha = dx^\alpha$ is also known as the holonomic 1-form. However, the exterior derivative of $e^a$ needs not to be zero, $de^a \neq 0$. Correspondingly, in the literature the coordinate indices are sometimes called as the holonomic indices and the orthonormal indices as the anholonomic indices. It should be noticed that in the coordinate frame $d(dx^\alpha)=0$, but $dg_{\alpha \beta} \neq 0$; in the orthonormal frame $de^a(x) \neq 0$, but $d\eta_{ab}=0$. Thus, apart from the coordinate and orthonormal basis it is always possible to work in a {\it mixed frame} in between them in which the exterior derivatives of both the metric components and the coframe are non-zero; $dg_{AB}(x) \neq 0$ and $de^A(x) \neq 0$.  Correspondingly, the metric given in the equation (\ref{metric1}) or (\ref{metric2}) turns out to be
  \begin{equation}\label{metric3}
   g = g_{AB}(x) e^A(x) \otimes e^B(x)
 \end{equation}
where $A, B, \cdots = \bar{0}, \bar{1}, \cdots , \bar{n}-\bar{1}$ denotes the mixed indices, see Table \ref{tab:clasification of frames}.
 \begin{table}[htbp]
 \caption{Classification of coordinate, orthonormal and mixed frames.}
 \centering
 \begin{tabular}{|c|c|c|}
 \hline
coordinate frame & orthonormal frame & mixed frame \\
 (holonomic)   & (Lorentzian, anholonomic) & (anholonomic)  \\
 \hline
  $dg_{\alpha \beta} \neq 0$   &$d\eta_{ab} = 0$ & $dg_{AB} \neq 0$\\
  \hline
  $de^\alpha = d^2x^\alpha =0$  & $de^a \neq 0$ & $de^A \neq 0$\\
  \hline
 \end{tabular}
 \label{tab:clasification of frames}
  \end{table}

We fix the orientation of the manifold by the Hodge map in the orthonormal coframe, $*1=\frac{1}{n!} \epsilon_{a_1a_2 \cdots a_n} e^{a_1} \wedge e^{a_2} \wedge \cdots \wedge e^{a_n} = e^0 \wedge e^1 \wedge \cdots \wedge e^{n-1}$ where $\wedge$ denotes the exterior product. Here $\epsilon_{a_1 a_2 \cdots a_n}$ with the choice $\epsilon_{01\cdots (n-1)}=+1$ denotes the totally antisymmetric Levi-Civita tensor. From now on we make use of the abbreviation $e^{ab\cdots} \equiv e^a \wedge e^b \wedge \cdots$. Another important operation in the exterior algebra is the interior product, $\iota_{X_a} \equiv \iota_a$ or $\iota_{\partial_\alpha} \equiv \iota_\alpha$, which maps a $p$-form to a $(p-1)$-form through the very basic relations
 \begin{equation}\label{duality3}
   \iota_a e^b = \delta^b_a \qquad \Leftrightarrow \qquad \iota_\alpha dx^\beta = \delta^\beta_\alpha.
 \end{equation}
These are another manifestations of the duality relations (\ref{duality2}) and (\ref{duality1}), respectively. Here again $\iota_a$ and $\iota_\alpha$ are related via the n-bein, $\iota_\alpha = h^a{}_\alpha \iota_a$. Additionally the interior product of any 0-form is zero by definition. It satisfies a very useful identity together with the Hodge map; $*(\psi \wedge e_a)=\iota_a *\psi$ where $\psi$ any $p$-form.

The affine connection $\nabla$ is determined by the affine connection 1-form ${\omega^a}_b$ by the relation $\nabla e^a = - \omega^a{}_b \wedge e^b$. Under the transition between the coordinate and the orthonormal frames defined in (\ref{doublet1}) or equivalently (\ref{doublet2}), for any $(p,q)$-type tensor-valued $r$-form $\mathfrak{T}^{a_1a_2 \cdots a_p}_{\; \; \; \; b_1b_2 \cdots b_q }$, $0 \leq p,q,r \leq n$,  to transform in a covariant way, .i.e., 
 \begin{align}
     \mathfrak{T}^{a_1a_2 \cdots a_p}_{\; \; \; \; b_1b_2 \cdots b_q } = h^{a_1}{}_{\alpha_1} \cdots h^{a_p}{}_{\alpha_p}  \mathfrak{T}^{\alpha_1 \alpha_2 \cdots \alpha_p}_{\; \; \; \; \beta_1 \beta_2 \cdots \beta_q } h^{\beta_1}{}_{b_1} \cdots h^{\beta_q}{}_{b_q}
 \end{align}
the affine connection 1-form must transform as
  \begin{equation}\label{doublet-con}
   \omega^a{}_b = h^a{}_\alpha \omega^\alpha{}_\beta h^\beta{}_b + h^a{}_\alpha dh^\alpha{}_b \qquad \Leftrightarrow \qquad
   \omega^\alpha{}_\beta = h^\alpha{}_a \omega^a{}_b h^b{}_\beta + h^\alpha{}_a dh^a{}_\beta .
 \end{equation}

On the other hand, unlike the transformation between the coordinate and orthonormal frames, there is a new transformation relating any two mixed coframes $e^A$ and $e^{A'}$ expressed by the formula
 \begin{align}
     e^{A'} = L^{A'}{}_A e^A \quad \text{or} \quad g_{A'B'} = L^A{}_{A'}g_{AB}L^B{}_{B'}
 \end{align}
where $L^{A'}{}_A, L^A{}_{A'} \in GL(n,\mathbb{R})$ and $L^{A'}{}_A L^A{}_{B'} = \delta^{A'}_{B'}$ and $L^{A}{}_{A'} L^{A'}{}_{B} = \delta^{A}_{B}$. Thus, we define the $GL(n,\mathbb{R})$-covariant exterior derivative of a $(p,q)$-type tensor-valued $r$-form $\mathfrak{T}^{A_1A_2 \cdots A_p}_{\; \; \; \; B_1B_2 \cdots B_q }$ as
 \begin{align}\label{covariant-derivative1}
   D \mathfrak{T}^{A_1A_2 \cdots A_p}_{\; \; \; \; B_1B_2 \cdots B_q } = d \mathfrak{T}^{A_1A_2 \cdots A_p}_{\; \; \; \; B_1B_2 \cdots B_q }
    &+ \omega^{A_1}{}_C \wedge \mathfrak{T}^{CA_2 \cdots A_p}_{\; \; \; \; B_1B_2 \cdots B_q } + \cdots + \omega^{A_p}{}_C \wedge \mathfrak{T}^{A_1A_2 \cdots C}_{\; \; \; \; B_1B_2 \cdots B_q } \nonumber \\
     &- \omega^C{}_{B_1} \wedge \mathfrak{T}^{A_1A_2 \cdots A_p}_{\; \; \; \; CB_2 \cdots B_q } - \cdots
    - \omega^C{}_{B_q} \wedge \mathfrak{T}^{A_1A_2 \cdots A_p}_{\; \; \; \; B_1B_2 \cdots C} .
 \end{align}
Now, we can state the Cartan structure equations for the nonmetricity tensor 1-form, the torsion tensor 2-form and the curvature tensor 2-form. They are written explicitly in the mixed frame, respectively, as follows
 \begin{subequations} \label{eq:cartan-mixedframe}
  \begin{align}
      Q_{AB} &:= - \frac{1}{2} D g_{AB}
                      = \frac{1}{2} (-d g_{AB} + \omega_{AB} + \omega_{BA}) , \label{nonmet}\\
     T^A &:= D e^A = d e^A + {\omega^A}_B \wedge e^B  ,\label{tors}\\
     {R^A}_B &:= D {\omega^A}_B := d {\omega^A}_B + {\omega^A}_C \wedge {\omega^C}_B ,  \label{curva}
  \end{align}
 \end{subequations}
where the factor $-1/2$ in the definition of nonmetricity is a convention\footnote{We choose it such in order to be able to write $\omega_{(ab)}=Q_{ab}$ via the equation (\ref{nonmet2}) where the round parenthesis in the subscript denotes the symmetry of enclosed indices, $\omega_{(ab)}=\frac{1}{2} (\omega_{ab} + \omega_{ba})$.}. They are not entirely independent because they satisfy the Bianchi identities
  \begin{align}
   D Q_{AB} = \frac{1}{2} ( R_{AB} +R_{BA}) ,  \qquad 
       D T^A  = {R^A}_B \wedge e^B , \qquad 
       D {R^A}_B = 0 . 
   \end{align}

Similar definitions and analysis can be repeated for the coordinate frame by replacing all capital Latin indices with little Greek indices. Then, only difference appears in the torsion among the Cartan structure equations because of $d(dx^\alpha)=d^2x^\alpha=0$,
 \begin{subequations} \label{eq:cartan-coordframe}
  \begin{align}
    Q_{\alpha \beta} &:= -\frac{1}{2} Dg_{\alpha \beta} = \frac{1}{2} (-d g_{\alpha \beta} + \omega_{\alpha \beta} + \omega_{\beta \alpha})  , \label{nonmet1}\\
    T^\alpha  &:= D e^\alpha = {\omega^\alpha}_\beta \wedge dx^\beta ,  \label{tors1}\\
    {R^\alpha}_\beta &:=  D {\omega^\alpha}_\beta := d {\omega^\alpha}_\beta + {\omega^\alpha}_\gamma \wedge {\omega^\gamma}_\beta .\label{curva1}
   \end{align}
 \end{subequations}
Again, the transformation elements of two coordinate frames form the general linear group: $dx^{\alpha'} = L^{\alpha'}{}_\alpha dx^\alpha$ or $g_{\alpha'\beta'} = L^\alpha{}_{\alpha'}g_{\alpha \beta}L^\beta{}_{\beta'}$ where $L^{\alpha'}{}_\alpha , L^\alpha{}_{\alpha'} \in GL(n, \mathbb{R})$. The Bianchi identities turn out to be 
  \begin{align}
   D Q_{\alpha \beta} = \frac{1}{2} ( R_{\alpha \beta} +R_{\beta \alpha}) ,  \qquad D T^\alpha   = {R^\alpha}_\beta \wedge dx^\beta , \qquad D {R^\alpha}_\beta = 0 . 
   \end{align}

When the similar steps are redone for the orthonormal frame by replacing capital Latin indices with little Latin indices, there appear a difference in the nonmetricity among the Cartan structure equations because of $d\eta_{ab}=0$,
   \begin{subequations} \label{eq:cartan-onframe}
 \begin{align}
  Q_{ab} &= -\frac{1}{2} D\eta_{ab} = \frac{1}{2} (\omega_{ab} + \omega_{ba}) , \label{nonmet2}\\
    T^a &:= De^a = d e^a + {\omega^a}_b \wedge e^b ,  \label{tors2}\\
    {R^a}_b &:=  D {\omega^a}_b := d {\omega^a}_b + {\omega^a}_c \wedge {\omega^c}_b .  \label{curva2}
   \end{align}
 \end{subequations}
But, this time the transformation elements of two orthonormal frames form the Lorentz group: $e^{a'} = L^{a'}{}_a e^a$ or $\eta_{a'b'} = L^a{}_{a'}\eta_{ab}L^b{}_{b'}$ where $L^{a'}{}_a , L^a{}_{a'} \in SO(1, n-1)$ because $ \eta_{a'b'} = \eta_{ab} = \text{diag}(-1, \underbrace{  1 , \cdots , 1}_{(n-1) \; many})$  \cite{frankel2012}.  This is the reason why $e^a$ is sometimes called as the Lorentzian coframe. The Bianchi identities take the below form
   \begin{align}
   D Q_{ab} = \frac{1}{2} ( R_{ab} +R_{ba}) ,  \qquad 
       D T^a  = {R^a}_b \wedge e^b , \qquad 
       D {R^a}_b = 0 . 
   \end{align}

In fact, it may be argued to use different symbols for covariant exterior derivatives because the transformation groups are different. However, we understand the correct covariant derivative by looking at the indices of tensor on which it acts.

\subsection{World of transformations}
\label{world}

It is worthy to give some remarks on transitions among the frames. In this paper there are six different transformations.
 \begin{enumerate}
    \item Between the orthonormal frame and the coordinate frame via n-bein $h^\alpha{}_a$ and its inverse
         \begin{subequations} \label{eq:orth-to-coordinate}
             \begin{align}
          e^a &= h^a{}_\alpha dx^\alpha , &  \eta_{ab} &= h^\alpha{}_a g_{\alpha \beta} h^\beta{}_b , &   \omega^a{}_b &= h^a{}_\alpha \omega^\alpha{}_\beta h^\beta{}_b + h^a{}_\alpha dh^\alpha{}_b , \label{eq:orth-to-coordinate-a}\\
           T^a &= h^a{}_\alpha T^\alpha , &  Q_{ab} &= h^\alpha{}_a Q_{\alpha \beta} h^\beta{}_b , &   R^a{}_b &= h^a{}_\alpha R^\alpha{}_\beta h^\beta{}_b , \label{eq:orth-to-coordinate-b}
            \end{align}
      \end{subequations}
      where $h^\alpha{}_a, h^a{}_\alpha \in GL(n, \mathbb{R})$. Since $e^a$ is $g$-orthonormal, $h^a{}_\alpha$  are determined by metric functions. Thus, we can write symbolically $h^a{}_\alpha = h^a{}_\alpha(g)$.
  \item Between the mixed frame and the coordinate frame via n-bein $h^\alpha{}_A$ and its inverse
      \begin{subequations}
        \begin{align}
          e^A &= h^A{}_\alpha dx^\alpha , &  g_{AB} &= h^\alpha{}_A g_{\alpha \beta} h^\beta{}_B , &   \omega^A{}_B &= h^A{}_\alpha \omega^\alpha{}_\beta h^\beta{}_B + h^A{}_\alpha dh^\alpha{}_B , \\
          T^A &= h^A{}_\alpha T^\alpha , &  Q_{AB} &= h^\alpha{}_A Q_{\alpha \beta} h^\beta{}_B , &   R^A{}_B &= h^A{}_\alpha R^\alpha{}_\beta h^\beta{}_B ,
          \end{align}
      \end{subequations}
      where $h^\alpha{}_A, h^A{}_\alpha \in GL(n, \mathbb{R})$. Since $e^A$ is determined from the $g$ metric, $h^A{}_\alpha$  are formed through metric functions. Thus, we can denote symbolically $h^A{}_\alpha = h^A{}_\alpha(g)$.
   \item Between the orthonormal frame and the mixed frame via n-bein $h^a{}_A$ and its inverse
      \begin{subequations} \label{eq:orth-to-mix}
         \begin{align}
          e^a &= h^a{}_A e^A , &  \eta_{ab} &= h^A{}_a g_{AB} h^B{}_b , &   \omega^a{}_b &= h^a{}_A \omega^A{}_B h^B{}_b + h^a{}_A dh^A{}_b , \label{eq:orth-to-mix-a}\\
           T^a &= h^a{}_A T^A , &  Q_{ab} &= h^A{}_a Q_{AB} h^B{}_b , &   R^a{}_b &= h^a{}_A R^A{}_B h^B{}_b , \label{eq:orth-to-mix-b}
         \end{align}
      \end{subequations}
      where $h^A{}_a, h^a{}_A \in GL(n, \mathbb{R})$. Since both $e^a$ and $e^A$ are determined by the $g$ metric, $h^a{}_A$  are made up from metric functions. Thus, we can write symbolically $h^A{}_a = h^A{}_a(g)$.
   \item Between two coordinate frames via transformation elements $L^{\alpha}{}_{\alpha'}$ and its inverse
       \begin{subequations}
          \begin{align}
          dx^{\alpha'} &= L^{\alpha'}{}_{\alpha} dx^\alpha , &  g_{\alpha' \beta'} &= L^\alpha{}_{\alpha'} g_{\alpha \beta} L^\beta{}_{\beta'} , &   \omega^{\alpha'}{}_{\beta'} &= L^{\alpha'}{}_{\alpha} \omega^\alpha{}_\beta L^\beta{}_{\beta'} + L^{\alpha'}{}_{\alpha} dL^{\alpha}{}_{\beta'} , \\
           T^{\alpha'} &= L^{\alpha'}{}_{\alpha} T^\alpha , &  Q_{\alpha' \beta'} &= L^\alpha{}_{\alpha'} Q_{\alpha \beta} L^\beta{}_{\beta'} , &   R^{\alpha'}{}_{\beta'} &= L^{\alpha'}{}_{\alpha} R^\alpha{}_\beta L^\beta{}_{\beta'} ,
          \end{align}
      \end{subequations}
      where $L^\alpha{}_{\alpha'}, L^{\alpha'}{}_\alpha \in GL(n, \mathbb{R})$. It is worthy to remark that transformation elements can be written in terms of a general coordinate transformation, $x^\alpha \to x^{\alpha'}\left(x^\alpha \right)$,
      as $L^{\alpha'}{}_\alpha = \partial x^{\alpha'}/\partial x^\alpha$ and $L^\alpha{}_{\alpha'} = \partial x^\alpha / \partial x^{\alpha'}$.
   \item Between two mixed frames via transformation elements $L^{A}{}_{A'}$ and its inverse
       \begin{subequations}
           \begin{align}
          e^{A'} &= L^{A'}{}_{A} e^A , &  g_{A' B'} &= L^A{}_{A'} g_{AB} L^B{}_{B'} , &   \omega^{A'}{}_{B'} &= L^{A'}{}_{A} \omega^A{}_B L^B{}_{B'} + L^{A'}{}_{A} dL^{A}{}_{B'} , \\
          T^{A'} &= L^{A'}{}_{A} T^A , &  Q_{A' B'} &= L^A{}_{A'} Q_{AB} L^B{}_{B'} , &   R^{A'}{}_{B'} &= L^{A'}{}_{A} R^A{}_B L^B{}_{B'} ,
         \end{align}
      \end{subequations}
      where $L^A{}_{A'}, L^{A'}{}_A \in GL(n, \mathbb{R})$.
  \item Between two orthonormal frames via transformation elements $L^{a}{}_{a'}$ and its inverse
       \begin{subequations} \label{eq:orth-fra-trans}
           \begin{align}
          e^{a'} &= L^{a'}{}_{a} e^a , &  \eta_{a' b'} &= L^a{}_{a'} \eta_{ab} L^b{}_{b'} , &   \omega^{a'}{}_{b'} &= L^{a'}{}_{a} \omega^a{}_b L^b{}_{b'} + L^{a'}{}_{a} dL^{a}{}_{b'} , \label{eq:eq:orth-fra-trans-a}\\
           T^{a'} &= L^{a'}{}_{a} T^a , &  Q_{a' b'} &= L^a{}_{a'} Q_{ab} L^b{}_{b'} , &   R^{a'}{}_{b'} &= L^{a'}{}_{a} R^a{}_b L^b{}_{b'} ,  \label{eq:eq:orth-fra-trans-b}
         \end{align}
      \end{subequations}
      where $L^a{}_{a'}, L^{a'}{}_a \in SO(1, n-1)$. Note that since $\eta_{a'b'}=\eta_{ab}=\text{diag}(-1, \underbrace{  1, 1, \cdots , 1}_{(n-1) \; many})$, the group formed by transformation elements is the Lorentz group rather than the general linear group \cite{frankel2012}. 
 \end{enumerate}

  \subsection{Decomposition of the full (affine) connection } \label{sec:connec-decomp}

In a {\it mixed frame} the full connection 1-form can be decomposed uniquely as follows \cite{benn1982,hehl1995,tucker1995},
 \begin{align}
     {\omega^A}_B =  \underbrace{ \frac{1}{2} g^{AF} (\iota_C dg_{FB} + \iota_B dg_{FC} - \iota_F dg_{BC})e^C + {\widetilde{\omega}^A}{}_B}_{Levi-Civita \;\,  or \;\,  Christoffel \;\,  or \; \, Riemannian}
     + \underbrace{  \underbrace{{K^A}_B}_{contortion} + \underbrace{{q^A}_B  + {Q^A}_B}_{disformation} }_{distortion}  \label{connect:dec}
 \end{align}
where ${\widetilde{\omega}}_{AB} = -{\widetilde{\omega}}_{BA}$ is the Levi-Civita connection 1-form
 \begin{align}
     {\widetilde{\omega}^A}{}_B \wedge e^B = -d e^A \quad \text{or} \quad \widetilde{\omega}_{AB} = \frac{1}{2} \left[ -\iota_A de_B + \iota_B de_A + (\iota_A \iota_B de_C) e^C \right]  , \label{LevCiv}
 \end{align}
$K_{AB} = -K_{BA}$ is the contortion tensor 1-form,
 \begin{align}
   {K^A}_B  \wedge e^B = T^A \quad \text{or} \quad K_{AB} = \frac{1}{2} \left[ \iota_A T_B - \iota_B T_A - (\iota_A \iota_B T_C) e^C \right]  ,  \label{contort}  
 \end{align}
and $q_{AB}$ are defined in terms of nonmetricity
 \begin{align}
    q_{AB} = -( \imath_A  Q_{BC} ) e^C + ( \imath_B Q_{AC}) e^C  . \label{q:ab}
 \end{align}
This decomposition is self-consistent. To see that it is enough to multiply (\ref{connect:dec}) from right by $\wedge e^B$ and to use the definitions above. While moving indices vertically in front of both $d$ and $D$, a special attention is needed because $d g_{AB} \neq 0$ and $D g_{AB} = -2 Q_{AB} \neq 0$. The symmetric part of the full connection comes from (\ref{nonmet})
 \begin{align}
  \omega_{(AB)} = Q_{AB } + \frac{1}{2} d g_{AB } \label{connect:sym}
 \end{align}
and the remainder is the anti-symmetric part
 \begin{align}
  \omega_{[AB]} = \frac{1}{2} (\iota_B dg_{AC} - \iota_A dg_{BC})e^C + \widetilde{\omega}_{AB} + K_{AB} + q_{AB} . \label{connect:ansym}
 \end{align}
If only $Q_{AB}=0$, the connection is said to be metric compatible. If both $Q_{AB}=0$ and $T^A =0$, the affine connection becomes the Riemannian connection. An affine geometry is classified whether nonmetricity, torsion and/or vanish or not, see Table  \ref{tab:clasification of spacetimes}.
   \begin{table}[ht]
\caption{Classification of spacetimes. In literature, sometimes firstly $Q_{AB}$ is decomposed as $Q_{AB}=\overline{Q}_{AB}+\frac{1}{n}g_{AB}Q$ where $g^{AB}Q_{AB}=Q$ and $g^{AB}\overline{Q}_{AB}=0$, then the case of $\overline{Q}_{AB}=0$ and $Q\neq 0$ is called Weyl geometry. But, here by ``Weyl geometry'' we mean $Q_{AB} \neq 0$ in general!}
 \centering
 \begin{tabular}{|c|c|c|l|}
 \hline
   $Q_{AB}$ & $T^A$ & $R^A{}_B$ & {\it Geometry Name}  \\
 \hline \hline
  $0$ & $0$ & $0$ & Minkowski  \\
  \hline
   $0$ & $ 0$ & $\neq 0$ & Riemann  \\
  \hline
  $0$ & $\neq 0$ & $ 0$ & Metric (Weitzenböck) teleparallel  \\
  \hline
  $\neq 0$ & $0$ & $0$ & Symmetric teleparallel  \\
  \hline
  $0$ & $\neq 0$ & $\neq 0$ & Riemann-Cartan  \\
  \hline
  $\neq 0$ & $0$ & $\neq 0$ & Riemann-Weyl \\
  \hline
   $\neq 0$ & $\neq 0$ & $ 0$ & General teleparallel  \\
  \hline
  $\neq 0$ & $\neq 0$ & $\neq 0$ & Riemann-Cartan-Weyl \\
  \hline
 \end{tabular}
 \label{tab:clasification of spacetimes}
  \end{table}

In the coordinate frame the decomposition (\ref{connect:dec}) reduces to
  \begin{equation}\label{decomp-con-coord}
{\omega^\alpha}_\beta =  \underbrace{ \frac{1}{2} g^{\alpha \sigma}(\partial_\gamma g_{\sigma \beta} + \partial_\beta g_{\sigma \gamma} - \partial_\sigma g_{\beta \gamma})dx^\gamma }_{Christoffel}
     + \underbrace{{K^\alpha}_\beta}_{contortion} + \underbrace{{q^\alpha}_\beta  + {Q^\alpha}_\beta}_{disformation}
 \end{equation}
where the first group on the right hand side is, in fact, the Christoffel symbols. In the orthonormal frame it takes the form
 \begin{align}
   \omega_{ab} = \underbrace{  \widetilde{\omega}_{ab}}_{Levi-Civita} + \underbrace{ K_{ab}}_{contortion} + \underbrace{ q_{ab}
          +  Q_{ab}}_{disformation} . \label{connect:on}    
 \end{align}
Here the Levi-Civita connection 1-form is 
 \begin{equation}\label{Levi-Civita:on}
   \widetilde{\omega}_{ab} = \frac{1}{2} \left[ -\iota_a de_b + \iota_b de_a + (\iota_a \iota_b de_c) e^c \right] \quad \text{or} \quad \widetilde{\omega}^a{}_b \wedge e^b = -de^a \;\; \text{with} \;\; \widetilde{\omega}_{ab} = - \widetilde{\omega}_{ba}
 \end{equation}
and the contortion 1-form is
   \begin{equation}\label{contorsion:on}
   K_{ab} = \frac{1}{2} \left[ \iota_a T_b - \iota_b T_a - (\iota_a \iota_b T_c) e^c \right] \quad \text{or} \quad K^a{}_b \wedge e^b = T^a \;\; \text{with} \;\; K_{ab} = - K_{ba} .
 \end{equation}
Besides the quantity $q_{ab}$ is defined in terms of the nonmetricity
 \begin{equation}\label{qab-orthnorm}
   q_{ab} = -( \imath_a  Q_{bc} ) e^c + ( \imath_b Q_{ac}) e^c . 
 \end{equation}

\subsection{Variational field equations}
\label{variational}

One way of obtaining field equations of a gravity theory is to vary the concerning Lagrangian with respect to independent variables. Thus, for a gravity model formulated in non-Riemannian spacetimes firstly a Lagrangian $n$-form is proposed in terms of $g_{AB}$, $e^A$, $\omega^A{}_B$, $Q_{AB}$, $T^A$, $R^A{}_B$ and a matter field $p$-form $\Psi$. Essentially, $\Psi$ represents all fields and quantities (except $g_{AB}$, $e^A$, $\omega^A{}_B$, $Q_{AB}$, $T^A$, $R^A{}_B$) such as scalar field, electromagnetic field, Lagrange multipliers, spinor field etc.  Meanwhile, since nonmetricity, torsion and curvature are defined in terms of metric, coframe and affine connection via the equations (\ref{eq:cartan-mixedframe}), the independent quantities are solely $g_{AB}$, $e^A$, $\omega^A{}_B$ and $\Psi$. Correspondingly, we denote any Lagrangian $n$-form as $L=L[g_{AB}, e^A, \omega^A{}_B, \Psi]$. Then, variations with respect to independent variables are computed,
 \begin{align}
     \delta L =& \delta g_{AB} \wedge \sigma^{AB}[g_{AB}, e^A, \omega^A{}_B, \Psi] + \delta e^A \wedge \tau_A[g_{AB}, e^A, \omega^A{}_B, \Psi] \nonumber \\ 
     &+ \delta \omega^A{}_B \wedge \Sigma^B{}_A[g_{AB}, e^A, \omega^A{}_B, \Psi] + \delta \Psi \wedge \mathcal{M}[g_{AB}, e^A, \omega^A{}_B, \Psi] + mod(d) ,     
 \end{align}
where $\sigma^{AB}=\sigma^{BA}$ are metric $n$-forms, $\tau_A$ are energy-momentum $(n-1)$-forms and $\Sigma^B{}_A$ are angular (hyper) momentum $(n-1)$-forms, $\mathcal{M}$ is matter $(n-p)$-form and $mod(d)$ is the exact form, 
\begin{align}
    mod(d) =&  d\big\{ \delta g_{AB} \wedge \mathcal{A}^{AB}[g_{AB}, e^A, \omega^A{}_B, \Psi] + \delta e^A \wedge \mathcal{B}_A[g_{AB}, e^A, \omega^A{}_B, \Psi] \nonumber \\
     & \qquad  + \delta \omega^A{}_B \wedge \mathcal{C}^B{}_A[g_{AB}, e^A, \omega^A{}_B, \Psi] + \delta \Psi \wedge \mathcal{F}[g_{AB}, e^A, \omega^A{}_B, \Psi]\big\} .
\end{align}
Here $\mathcal{A}^{AB}=\mathcal{A}^{BA}$ are some $(n-1)$-forms,  $\mathcal{B}_A$ are some $(n-2)$-forms, $\mathcal{C}^B{}_A$ are some $(n-2)$-forms and $\mathcal{F}$ is a $(n-p-1)$-form. It is a worthy remark that exact form $mod(d)$ does not contribute to the variational field equations, but it can be useful in calculation of the Noether charges \cite{BeltranJimenez:2021kpj}.  Accordingly, the Hamilton principle, $\delta L=0$, yields the field equations, 
  \begin{subequations}
 \begin{align}
     \sigma^{AB}[g_{AB}, e^A, \omega^A{}_B, \Psi] &=0, \qquad \textsc{metric eqn} \\
     \tau_A[g_{AB}, e^A, \omega^A{}_B, \Psi] &=0 , \qquad \textsc{coframe eqn} \\
      \Sigma^B{}_A[g_{AB}, e^A, \omega^A{}_B, \Psi] &=0 , \qquad \textsc{connection eqn} \\ 
      \mathcal{M}[g_{AB}, e^A, \omega^A{}_B, \Psi] &=0 . \qquad \, \textsc{matter eqn} 
 \end{align}
 \end{subequations}
In this derivation since we use the mixed frame, the Lagrangian is invariant under $GL(n, \mathbb{R})$ transformations. Lorentz invariance and diffeomorphism invariance of the theory are analyzed separately and then it is seen that metric, coframe and connection equations are not independent. Similar formulations and remarks are valid for the the coordinate frame.

However, when we formulate the theory in terms of orthonormal exterior forms from the outset, since it is to be the coordinate independent, the Lagrangian is invariant under a diffeomorphism transformation by construction. Besides, since in an orthonormal basis the metric components are fixed, $\delta \eta_{ab}=0$, the variation of $L[\eta_{ab}, e^a, \omega^a{}_b, \Psi]$ with respect to the metric is equivalently accomplished by variation with respect to the orthonormal basis 1-form, $e^a$. Thus, we obtain
  \begin{align} \label{eq:var-lagr-orthonormal1}
     \delta L =& \, \delta e^a \wedge \tau_a[\eta_{ab}, e^a, \omega^a{}_b, \Psi]  
     + \delta \omega^a{}_b \wedge \Sigma^b{}_a[\eta_{ab}, e^a, \omega^a{}_b, \Psi] \nonumber \\
     & \qquad \qquad \qquad \qquad \qquad \qquad \qquad + \delta \Psi \wedge \mathcal{M}[\eta_{ab}, e^a, \omega^a{}_b, \Psi] + mod(d),
 \end{align}
and then the variational field equations turn out to be
  \begin{subequations} \label{eq:var-field-eqns-orth}
 \begin{align}
     \tau_a[\eta_{ab}, e^a, \omega^a{}_b, \Psi] &=0 , \qquad \textsc{coframe eqn} \label{eq:coframe-eqn-orth}\\
      \Sigma^b{}_a[\eta_{ab}, e^a, \omega^a{}_b, \Psi] &=0 , \qquad \textsc{connection eqn} \label{eq:connection-eqn-orth}\\ 
      \mathcal{M}[\eta_{ab}, e^a, \omega^a{}_b, \Psi] &=0. \qquad \textsc{matter eqn} \label{eq:matter-eqn-orth}
 \end{align}
 \end{subequations}
Here, in fact, the coframe and connection equations are not totally independent as well. Let us make it clear by counting the components of equations and the unknowns. Firstly the number of components of any $p$-form is obtained by $\frac{n!}{p! (n-p)!}$. So, the coframe equation $(n-1)$-form yields $n^2$ many equations and  the connection equation $(n-1)$-form $n^3$ many equations adding up $n^2(n+1)$ in total. On the other hand, the number of unknowns coming from the orthonormal coframe is $n(n+1)/2$ because $e^a$ is metric-orthonormal and $n^3$ coming from the full connection 1-form, $\omega^a{}_b$. They give the total number of unknowns as $n^3+n(n+1)/2$. Let us assume that the components (unknowns) of matter field is equal to the number of components of the matter equation. Consequently, the number of equations is larger than the number of unknowns by $n(n-1)/2$. Now, we want to explain this discrepancy. To ensure the independence of variational equations from the choice of orthonormal basis, $e^a$, the Lagrangian must be invariant under $SO(1,n-1)$ transformations. We can see it by taking the variations of $e^a$ and $\omega^a{}_b$ under an infinitesimal Lorentz transformation, $\varepsilon^a{}_b(x)$. Let us rewrite the transformation rules given by (\ref{eq:eq:orth-fra-trans-a}) in a slightly different but more convenient notation,
   \begin{align}
       \overset{\prime}{e}{}^a = L^a{}_b e^b  \qquad \text{and} \qquad \overset{\prime}{\omega}{}^a{}_b = L^a{}_c \omega^c{}_d \left(L^{-1}\right)^d{}_b + L^a{}_c d\left(L^{-1}\right)^c{}_b,
 \end{align}
where $L^a{}_b(x) = \delta^a_b + \varepsilon^a{}_b(x)$ and $\left(L^{-1}\right)^a{}_b= \delta^a_b - \varepsilon^a{}_b(x)$ such that $\varepsilon_{ab}=-\varepsilon_{ba}$. Then, the variations of $e^a$ and $\omega^a{}_b$ cause to
  \begin{subequations}
       \begin{align}
       \delta e^a &= \overset{\prime}{e}{}^a - e^a = \varepsilon^a{}_b e^b ,\\
        \delta \omega^a{}_b &= \overset{\prime}{\omega}{}^a{}_b - \omega^a{}_b = - D \varepsilon^a{}_b,
  \end{align}
  \end{subequations}
where $ D \varepsilon^a{}_b := d\varepsilon^a{}_b + \omega^a{}_c \varepsilon^c{}_b - \omega^c{}_b \varepsilon^a{}_c$. By substituting these two results into the equation (\ref{eq:var-lagr-orthonormal1}) we arrive at
   \begin{subequations}\label{eq:var-lagr-orthonormal2}
     \begin{align} 
     \delta L =& \, \varepsilon^a{}_b e^b \wedge \tau_a[\eta_{ab}, e^a, \omega^a{}_b, \Psi]  
     -  D\varepsilon^a{}_b \wedge \Sigma^b{}_a[\eta_{ab}, e^a, \omega^a{}_b, \Psi] \nonumber \\
     & \qquad \qquad \qquad \qquad \qquad \qquad \qquad + \delta \Psi \wedge \mathcal{M}[\eta_{ab}, e^a, \omega^a{}_b, \Psi] + mod(d) \label{eq:var-lorentz-a} \\
      =& \, \varepsilon^a{}_b e^b \wedge \tau_a[\eta_{ab}, e^a, \omega^a{}_b, \Psi]  
     + \varepsilon^a{}_b \wedge D\Sigma^b{}_a[\eta_{ab}, e^a, \omega^a{}_b, \Psi] \nonumber \\
     & \qquad \qquad \qquad \qquad \qquad \qquad \qquad  + \delta \Psi \wedge \mathcal{M}[\eta_{ab}, e^a, \omega^a{}_b, \Psi] + mod(d) .  \label{eq:var-lorentz-b}
 \end{align}
 \end{subequations}
While passing from (\ref{eq:var-lorentz-a}) to (\ref{eq:var-lorentz-b}) we used 
 \begin{align}
     D\varepsilon^a{}_b \wedge \Sigma^b{}_a =  d(\varepsilon^a{}_b \Sigma^b{}_a) - \varepsilon^a{}_b D\Sigma^b{}_a
 \end{align}
and put the exact form, $d(\varepsilon^a{}_b \Sigma^b{}_a)$, inside $mod(d)$. In order to be able to use the anti-symmetry property of $\varepsilon^{ab}$ we have to lower the index $b$ inside $D\Sigma^b{}_a$ in the equation (\ref{eq:var-lorentz-b}),
   \begin{align} 
     \delta L =& \, \varepsilon^{ab} \big\{ e_b \wedge \tau_a[\eta_{ab}, e^a, \omega^a{}_b, \Psi]  
     + 2 Q^{c}{}_b \wedge \Sigma_{ca} +  D\Sigma_{ba}[\eta_{ab}, e^a, \omega^a{}_b, \Psi] \big\} \nonumber \\
     & \qquad \qquad \qquad \qquad \qquad \qquad \qquad + \delta \Psi \wedge \mathcal{M}[\eta_{ab}, e^a, \omega^a{}_b, \Psi] + mod(d) .  
 \end{align}
When $Q_{ab} \neq 0$, lowering or raising an index in  front of $D$ is not trivial because of $D\eta_{ab}=-2Q_{ab}$ and $D\eta^{ab}= 2Q^{ab}$. Thus, since $\varepsilon^{ab}= -\varepsilon^{ba}$ can be taken to be arbitrary at each point, $\delta L=0$ gives 
  \begin{align}
      &D\Sigma_{[ab]}[\eta_{ab}, e^a, \omega^a{}_b, \Psi] + e_{[a} \wedge \tau_{b]}[\eta_{ab}, e^a, \omega^a{}_b, \Psi]  \nonumber \\
       & \qquad \qquad +  Q^{c}{}_a \wedge \Sigma_{cb}[\eta_{ab}, e^a, \omega^a{}_b, \Psi] - Q^{c}{}_b \wedge \Sigma_{ca}[\eta_{ab}, e^a, \omega^a{}_b, \Psi] =0
  \end{align}
apart from the matter equation (\ref{eq:matter-eqn-orth}). Here, the square bracket in the subscript indicates the anti-symmetry of enclosed indices, $e_{[a} \wedge \tau_{b]} = \frac{1}{2} (e_a \wedge \tau_b - e_b \wedge \tau_a)$. This result expresses the fact that the coframe equation (\ref{eq:coframe-eqn-orth}) and connection equation (\ref{eq:connection-eqn-orth}) are not all independent and their number is reduced by $n(n-1)/2$ which fixes the discrepancy between the numbers of unknowns and field equations counted above.

As a final note we want to say that when there is a Hodge star in Lagrangian it is not a straightforward to vary it. In those cases, we will perform calculation of variation by using the generic result from the Ref.\cite{adak2006},
 \begin{align}
      \delta(\alpha \wedge *\beta) = \delta \alpha \wedge *\beta + \delta \beta \wedge *\alpha - \delta e^a \wedge \big[ (\iota_a\beta) \wedge *\alpha - (-1)^p \alpha \wedge (\iota_a *\beta) \big] 
 \end{align}
where $\alpha$ and $\beta$ are some two $p$-forms in $n$ dimensions, ($0 \leq p \leq n$).

\subsection{Decomposition of the full (non-Riemannian) curvature}

It is sometimes useful to write the affine connection as Riemannian plus non-Riemannian parts, $\omega_{ab} = \widetilde{\omega}_{ab} + N_{ab}$ where $N_{ab} := K_{ab} + q_{ab} + Q_{ab}$ is called the distortion tensor 1-form.  Accordingly, the full curvature 2-form can be split into Riemannian plus non-Riemannian pieces
 \begin{align}
     R^a{}_b = \widetilde{R}^a{}_b + \widetilde{D}N^a{}_b + N^a{}_c \wedge N^c{}_b
 \end{align}
where $\widetilde{R}^a{}_b$ are the Riemannian curvature 2-form and $\widetilde{D}N^a{}_b$ is the covariant exterior derivative of $N^a{}_b$ with respect to the Levi-Civita connection,
 \begin{subequations}
     \begin{align}
         \widetilde{R}^a{}_b &= \widetilde{\omega}^a{}_{b} + \widetilde{\omega}^a{}_{c} \wedge \widetilde{\omega}^c{}_{b} , \\
         \widetilde{D}N^a{}_b &= dN^a{}_b +  \widetilde{\omega}^a{}_{c} \wedge N^c{}_b - \widetilde{\omega}^c{}_{b} \wedge N^a{}_c .
     \end{align}
 \end{subequations}
We always put a tilde sign over a Riemannian quantity throughout this paper. It is common to decompose the Einstein-Hilbert $n$-form as well
  \begin{align}
       R^a{}_b \wedge *e_a{}^b = \widetilde{R}^a{}_b \wedge *e_a{}^b + N^a{}_c \wedge N^c{}_b \wedge *e_a{}^b +  d \left( N^a{}_b \wedge *e_a{}^b \right)
  \end{align}
where $\widetilde{D} *e_a{}^b =0$ is valid. Since the last term is exact, it does not contribute to the variational field equations and therefore dismissed.

Similar decomposition can be written readily for the coordinate and mixed frames as well. In literature, mainly the coordinate frame and the orthonormal frame are used. The mixed frame is seldom preferred in explicit calculations since such can be facilitated by a suitable gauge-fixing, but the mixed frame is useful in theoretical considerations of gravity \cite{hehl1995,BeltranJimenez:2021kpj}. Let us also  reiterate that besides gravity, similar calculations are pursued in various rather different contexts. To highlight an interesting example, non-Riemannian geometry is relevant in the description of crystal defects \cite{roychowdhury-gupta-2017}. %Nonetheless, as far as we know this is the first work in literature in which the mixed frame is used for explicit calculations of an alternative theory of gravity.

In the calculations the following identities will be useful,
  \begin{align}
    D*e_{a_1} &= - Q \wedge *e_{a_1} + * e_{a_1a_2} \wedge T^{a_2} , \nonumber \\
    D*e_{a_1a_2} &= -  Q \wedge *e_{a_1a_2} + * e_{a_1a_2a_3} \wedge T^{a_3} , \nonumber \\
   \vdots & \label{ozedeslik D hodge kocerceve} \\
 D*e_{a_1 a_2 \cdots a_{n-1}}  &= -  Q \wedge *e_{a_1 a_2 \cdots a_{n-1}} + e_{a_1 a_2 \cdots a_{n-1}a_n} \wedge T^{a_n} , \nonumber \\
   D*e_{a_1 a_2 \cdots a_n}  &= -  Q \wedge *e_{a_1 a_2 \cdots a_n} , \nonumber \\
    D\eta_{ab} &= -2Q_{ab} \, , \quad D\eta^{ab} = +2Q^{ab} \, , \quad D\delta^a_b =0 , \nonumber
   \end{align}
where $ Q :=  \eta^{ab} Q_{ab} =  Q^a{}_a =\omega^a{}_a$ is the trace 1-form of nonmetricity.

\section{Metric formulation of teleparallel geometries} \label{sec:metric-formulation-teleparallel-geometries}

We will show how one can construct any of the three types of teleparallel geometries solely from a given metric tensor. Although it is done analytically in Riemannian geometry, we arrive at our construction here by exploiting gauge freedoms in a non-analytical way. 

\subsection{Riemannian geometry and Einstein's theory of gravity}

Since the Einstein's theory of gravity, general relativity, is accommodated in the Riemannian spacetime, features of this geometry are very well known. Therefore, we summarize them very briefly as reference for the teleparallel geometries and modified theories of gravity developed on them. We start in the orthonormal frame and apply the constraints $Q_{ab}=0$, $T^a=0$, $R^a{}_b \neq 0$ in the Cartan equations (\ref{eq:cartan-onframe}). The first two equations yield algebraic relations for $\omega^a{}_b$ and can be solved analytically. It is called the Levi-Civita connection 1-form, $\omega_{ab}=\widetilde{\omega}_{ab}$ and given by the equation (\ref{Levi-Civita:on}). Consequently we obtain all quantities from just metric functions; $g \to e^a(g) \to \widetilde{\omega}^a{}_b(g) \to \widetilde{R}^a{}_b(g)$.

The Einstein's theory of gravity is represented by the following Lagrangian $n$-form
 \begin{align}
     L_{GR} = \widetilde{L}_{EH} + \frac{\Lambda}{\kappa} *1 - L[mat]  
     % + T^a \wedge \lambda_a  +  Q_{ab} \wedge \alpha^{ab} 
 \end{align}
where $\kappa$ is a coupling constant, %$\lambda_a$ is the Lagrange multiplier $(n-2)$-form constraining torsion to zero, $\alpha^{ab}=\alpha^{ba}$ is the Lagrange multiplier $(n-1)$-form constraining the connection to be metric compatible, 
$\Lambda$ is the cosmological constant and $\widetilde{L}_{EH}$ is the Einstein-Hilbert Lagrangian,
 \begin{align*}
     \widetilde{L}_{EH} = \frac{1}{2\kappa} \widetilde{R}^a{}_b \wedge *e_a{}^b .
 \end{align*}
 %As long as $L[mat]$ is independent of the connection, the variation with respect to it gives $\lambda_a=0$ and $\alpha^{ab}=0$. 
 Thus, the variation with respect to the orthonormal coframe yields the Einstein's equation
 \begin{align} \label{eq:einsteineqn}
     -\frac{1}{2}\widetilde{R}^b{}_c \wedge *e_{ab}{}^c + \Lambda *e_a = \kappa \tau_a[mat]
 \end{align}
where $\tau_a[mat]$ is the energy-momentum $(n-1)$-form derived from $L[mat]$ via $\delta L[mat]= \delta e^a \wedge \tau_a[mat]$. It is worthwhile to remark two points. (i) For $n=2$, $\widetilde{R}^a{}_b$ has only one component and $L_{EH}$ is an exact form. Thus general relativity is trivial in two dimensions. (ii) For $n=3$ and vacuum, i.e., $L[mat]=0$, with $\Lambda=0$, there are no propagating (or dynamical) degrees of freedom of $\widetilde{R}^a{}_b$. That is because the number of independent components of the metric is $n(n+1)/2$, but the diffeomorphism symmetry eliminates $2n$ dynamical modes leaving $n(n-3)/2$. Since in four and higher dimensions this number is positive, Einstein's theory of gravity predicts propagating modes in vacuum.

\subsection{Metric (Weitzenböck) teleparallel geometry and gravity}

We start in the orthonormal frame and apply the constraints $Q_{ab}=0$, $R^a{}_b = 0$ whilst $T^a \neq 0$ in the Cartan equations (\ref{eq:cartan-onframe}). Because of the third equation, there is no direct algebraic relation for the components $\omega_{ab}$, and they can not be solved analytically. So, following the standard method one would make an ansatz for the metric and the connection independently, and then check if they satisfy the constraints above. There is an alternative method which we will explain now.

Let $O$ and $O'$ be two Lorentzian observers in this spacetime,
  \begin{subequations}
       \begin{align}
     O  &: \qquad e^a \qquad \, \text{and} \qquad \omega^a{}_b \\
     O' &: \qquad e^{a'} \qquad \text{and} \qquad \omega^{a'}{}_{b'}
      \end{align}
  \end{subequations}
The observer $O$ chooses the gauge potential (affine connection) as $\omega^a{}_b=0$ and arrives at the configuration 
      \begin{align}
   \omega_{ab}=0 \quad \Rightarrow \quad Q_{ab}=0, \quad R^a{}_b =0, \quad T^a = de^a \neq 0 . \label{eq:wtpg}
  \end{align}
In this case, via Eqn.(\ref{eq:orth-fra-trans}) the observer $O'$ reads the inertial connection $\omega^{a'}{}_{b'} = L^{a'}{}_a dL^{a}{}_{b'}$  corresponding to Eqn.(2) of Ref.\cite{koivisto-jimenez-2020} and the Minkowski metric $\eta_{a'b'} = L^a{}_{a'} \eta_{ab}L^a{}_{b'}$ corresponding to Eqn.(22) of Ref.\cite{koivisto-jimenez-2020}. As a complementary remark the primed observer reads the Cartan tensors through (\ref{eq:eq:orth-fra-trans-b}). Consequently, we obtain all quantities from just metric functions: for the unprimed observer $g \to e^a(g) \to T^a(g)$ together with ${\omega}^a{}_b =0$ and for the primed observer  $g \to e^{a'}(g)  \to L^{a'}{}_{a}(g) \to \omega^{a'}{}_{b'}(g)   \to T^{a'}(g)$.

From a different perspective, the choice $\omega^a{}_b=0$ may appear opposite to the spirit of relativity theory as we then seem to propose a set of connection components in a special frame. However, as long as we adhere to the conventional description of gravity as dynamical spacetime geometry (Riemannian or otherwise), the choice of the connection reflects a mere gauge redundancy and there is nothing special in the so called Weitzenb\"ock frame $\omega^a{}_b=0$. Only in the properly relativistic theory of gravity, the distinction can be made between ``inertial'' and ``non-inertial'' frames. This may be possible if the precise meaning of an ``inertial frame'' is that the Noether charges match with the observables, since the robust definition of the conserved charges is sensitive to the reference connection \cite{Koivisto:2022nar,BeltranJimenez:2021kpj}. Even then, the connection 
$\omega^a{}_b$ by itself has no physical significance, but what matters is how this connection is adjusted with respect to the coframe that one has taken to describe the situation at hand. 

The theory of Weitzenböck (metric) gravity is represented by the total Lagrangian 
  \begin{align} \label{eq:wtpg-lag}
     L_{WTP} = L_{T^2} + \Lambda *1  -  L[mat] +  Q_{ab} \wedge \alpha^{ab} +   R^a{}_b \wedge \rho^b{}_a  
 \end{align}
where the torsion squared even parity Lagrangian is
 \begin{align} 
     L_{T^2} =  k_1  T^a \wedge *T_a + k_2 \left(T^a \wedge e_a\right) \wedge *\left(T^b \wedge e_b\right) + k_3 (T^a \wedge e_b) \wedge *(T^b \wedge e_a) . \label{eq:torsionsquarelagrangian}
 \end{align}
Here $k_1, k_2, k_3$ are coupling constants, $\alpha^{ab}=\alpha^{ba}$ is a Lagrange multiplier $(n-1)$-form constraining the nonmetricity to zero, $\rho^b{}_a$ is Lagrange multiplier $(n-2)$-form constraining the full curvature to zero. Variations with respect to $e^a, \omega^a{}_b, \alpha^{ab}, \rho^b{}_a$ yield the field equations of Weitzenböck teleparallel gravity, respectively,
 \begin{subequations}
  \begin{align}
       \tau_a[T] + \Lambda *e_a  &= \tau_a[mat], & &\textsc{coframe eqn}  \label{eq:wtpg-field-eqn} \\
     \Sigma^b{}_a[T] + \alpha^b{}_a + D\rho^b{}_a &= \Sigma^b{}_a[mat], & &\textsc{connection eqn}   \label{eq:wtpg-field-eqn2} \\
     Q_{ab}&=0 ,  & &\textsc{metricity eqn} \label{zeroQ} \\
     R^a{}_b &=0 , & &\textsc{zero-torsion eqn} \label{zeroR}
 \end{align}
  \end{subequations}
where $\tau_a[mat]$ is matter energy-momentum and $\Sigma^b{}_a[mat]$ is matter angular momentum $(n-1)$-forms derived from $L[mat]$ via the relation $\delta L[mat]= \delta e^a \wedge \tau_a[mat] + \delta \omega^a{}_b \wedge \Sigma^b{}_a[mat]$, 
 \begin{align}
   \tau_a[T] = \sum_{i=1}^3 k_i \overset{(i)}{\tau_a}[T]  \qquad \text{and} \qquad
   \Sigma^b{}_a[T] = \sum_{i=1}^3 k_i \overset{(i)}{\Sigma^b{}_a}[T]
  \end{align}
together with
 \begin{subequations} \label{eq:enr-mom-tors-sqred}
 \begin{align}
    \overset{(1)}{\tau_a}[T] =& 2 D * T_a  - \left( \iota_a  T^b  \right) \wedge * T_b + T_b \wedge \left( \iota_a* T^b  \right)  , \\
    \overset{(2)}{\tau_a}[T] =& 2 D \left[ e_a \wedge *\left( T^b \wedge e_b \right)\right]  + 2T_a \wedge *\left( T^b \wedge e_b \right) - \left[ \iota_a\left( T^b \wedge e_b \right) \right] \wedge *\left( T^c \wedge e_c \right) \nonumber \\
    &- \left( T^c \wedge e_c \right) \wedge \left[ \iota_a*\left( T^b \wedge e_b \right)\right]  , \\
    \overset{(3)}{\tau_a}[T] =& 2 D \left[ e_b \wedge *\left( T^b \wedge e_a \right)\right]  + 2T^b \wedge *\left( T_a \wedge e_b \right) - \left[ \iota_a\left( T^b \wedge e_c \right) \right] \wedge *\left( T^c \wedge e_b \right) \nonumber \\
    &- \left( T^c \wedge e_b \right) \wedge \left[ \iota_a*\left( T^b \wedge e_c \right)\right]  ,
 \end{align}
 \end{subequations}
and 
  \begin{subequations} \label{eq:ang-mom-tors-sqred}
  \begin{align}
       \overset{(1)}{\Sigma^b{}_a}[T] =& 2 e^b \wedge *T_a  ,\\
      \overset{(2)}{\Sigma^b{}_a}[T] =& 2 e^b \wedge e_a \wedge *\left( T_c \wedge e^c \right) ,\\
      \overset{(3)}{\Sigma^b{}_a}[T] =& 2 e^b \wedge e^c \wedge *\left( T_c \wedge e_a \right) .
  \end{align}
   \end{subequations}
While dynamics of gravity is governed by the equation (\ref{eq:wtpg-field-eqn}), the other equation (\ref{eq:wtpg-field-eqn2}) is used for determining the Lagrange multipliers. The configuration below is the GR-equivalent values (for $n \geq 3$)
 \begin{align}
     k_1=0 , \qquad k_2=\frac{1}{4\kappa} , \qquad k_3=-\frac{1}{2\kappa} .
 \end{align}
In two dimensions the second and the third terms in the Lagrangian (\ref{eq:torsionsquarelagrangian}) disappear, because they contain 3-forms.
The propagating degrees of freedom have been checked for the generic Lagrangian in $n$ dimensions \cite{Koivisto:2018loq}. The number of independent components of the fields in metric teleparallelism in $n^2$. This is immediately seen in the orthonormal frame by imposing the Weizenb\"ock condition, such that all the components of the $n$-bein (and only those) are independent. Alternatively, in the coordinate frame, the symmetric pieces of the $n$-dimensional square matrix encoded into the metric are complemented by the antisymmetric pieces encoded into the flat and metric connection generated by a Lorentz transformation, and the sum is of course again $n^2$. General relativity does not propagate local degrees of freedom in $n=2$ or $n=3$, but the conclusion can be different for more generic metric teleparallel Lagrangians. Note also that the number of field equations is always equal to the number of independent components, though now this might not be so obvious. Implementing (\ref{zeroQ}) and (\ref{zeroR}) leaves us the metric-compatible pure-gauge connection, for which the equation is now the covariant derivative of (\ref{eq:wtpg-field-eqn2}). This covariant derivative
of the equation has then $n(n-1)/2$ free components. The remaining $n(n+1)/2$ independent equations are (\ref{eq:wtpg-field-eqn}). Though this coframe equation of motion appears to include too many components, we recall from the clarification in Section \ref{variational} that the antisymmetric components of the equation are not independent but degenerate with (\ref{eq:wtpg-field-eqn2}).

%We make similar remarks with above. (i) In two dimensions the second and third terms in the Lagrangian (\ref{eq:torsionsquarelagrangian}) disappears, because they contain 3-forms. While the number of independent components of torsion is two, that of the field equation is four. That is, like in general relativity, there is not any dynamical degree of freedom here too. (ii) In three dimensions the numbers of independent components of both torsion and the field equation are nine. Therefore, like in general relativity, there is not any propagating degree of freedom in vacuum. There are papers discussing the Weitzenböck teleparallel (metric teleparallel) gravity by following the steps defined above, e.g., \cite{yakov-itin-1999}.

\subsection{Symmetric teleparallel geometry and gravity}

If we start in the orthonormal frame by following the steps in the metric teleparallel geometry, in symmetric teleparallelism we can not proceed so easily. However, if we start this time in the coordinate frame and apply the constraints $T^\alpha = 0$, $R^\alpha{}_\beta = 0$ whilst $Q_{\alpha \beta} \neq 0$ in the Cartan equations (\ref{eq:cartan-coordframe}), we easily achieve the desired result. Again, because of the third equation, ${\omega^\alpha}_\beta$ can not be solved analytically. General strategy is to make ansatz for metric and connection independently, then checks if they satisfy the constraints above. Instead of that, one alternative path to be traced is to choose a very convenient coordinate system which can be seen as a gauge fixing such that 
 \begin{align}
   {\omega^\alpha}_\beta = 0 \qquad \Rightarrow \qquad Q_{\alpha \beta} = - \frac{1}{2} d g_{\alpha \beta} \neq 0, \qquad T^\alpha =0 , \qquad  {R^\alpha}_\beta =0 . \label{eq:stpg}
 \end{align}
This choice is called the {\it natural gauge or coincident gauge} \cite{adak-2006-tjp,tomi-lavinia2017}. Since the variational field equations are expressed in the orthonormal frame, we have to obtain the corresponding quantities by substituting the findings (\ref{eq:stpg}) into (\ref{eq:orth-to-coordinate}). 

Now let us repeat the result in the reverse order. The teleparallelism condition $R^a{}_b =0$ alone is satisfied by the inertial connection $\omega^a{}_b = h^a{}_\alpha dh^\alpha{}_b$ where $h^\alpha{}_a \in GL(n,\mathbb{R})$. If we also want to reset the torsion, we must write the coframe as $e^a = h^a{}_\alpha dx^\alpha$ meaning $h^a{}_\alpha =h^a{}_\alpha (g)$. Then, we arrive at $Q_{ab}=- \frac{1}{2} h^\alpha{}_a h^\beta{}_b dg_{\alpha \beta} \neq 0$.  Consequently, we obtain all quantities from just metric functions; $g \to e^a(g) \to h^a{}_\alpha(g) \to {\omega}^a{}_b(g) \to Q_{ab}(g)$.

The theory of symmetric teleparallel gravity is represented by the total Lagrangian 
  \begin{align} \label{eq:lag-stpg}
     L_{STP} = L_{Q^2} + \Lambda *1 -  L[mat] +  T^a \wedge \lambda_a +  R^a{}_b \wedge \rho^b{}_a  
 \end{align}
where the nonmetricity squared even parity Lagrangian is
 \begin{align} 
     L_{Q^2} =&  c_1  Q_{ab} \wedge *Q^{ab} + c_2 \left(Q_{ab} \wedge e^b\right) \wedge *\left(Q^{ac} \wedge e_c\right) + c_3 (Q_{ab} \wedge e_c) \wedge *(Q^{ac} \wedge e^b)  \nonumber \\ 
     &+ c_4 Q \wedge *Q   + c_5 \left(Q \wedge e^b\right) \wedge *\left(Q_{ab} \wedge e^a\right) . \label{eq:nonmetricitysquaredlagrangian}
 \end{align}
Here $c_i$, $i=1,2, \cdots ,5$, are coupling constants, $\lambda_a$ is Lagrange multiplier $(n-2)$-form constraining torsion to zero, $\rho^b{}_a$ is Lagrange multiplier $(n-2)$-form constraining the full curvature to zero. Variations with respect to $e^a, \omega^a{}_b, \lambda_a, \rho^b{}_a$ give the field equations of symmetric teleparallel gravity, respectively,
  \begin{subequations}
  \begin{align}
       \tau_a[Q] + \Lambda *e_a + D\lambda_a  &= \tau_a[mat]  , & &\textsc{coframe eqn} \label{eq:stpg-field-eqn1} \\
     \Sigma^b{}_a[Q] + e^b \wedge \lambda_a + D\rho^b{}_a &= \Sigma^b{}_a[mat]  , & &\textsc{connection eqn} \label{eq:stpg-field-eqn2} \\
     T^a &=0 ,  & &\textsc{zero-torsion eqn} \label{zerot} \\
     R^a{}_b &=0 , & &\textsc{zero-curvature eqn} \label{zeror}
 \end{align}
  \end{subequations}
where $\tau_a[mat]$ is matter energy-momentum and $\Sigma^b{}_a[mat]$ is matter angular momentum $(n-1)$-forms derived from $L[mat]$ via the relation $\delta L[mat]= \delta e^a \wedge \tau_a[mat] + \delta \omega^a{}_b \wedge \Sigma^b{}_a[mat]$, 
 \begin{align} 
 \tau_a[Q]= \sum_{i=1}^5 c_i \overset{(i)}{\tau_a}[Q] \qquad \text{and} \qquad 
 \Sigma^b{}_a[Q]= \sum_{i=1}^5 c_i \overset{(i)}{\Sigma^b{}_a}[Q]
 \end{align}
together with
  \begin{subequations} \label{eq:enr-mom-nonmet-sqred}
 \begin{align}
    \overset{(1)}{\tau_a}[Q] =& - \left( \iota_a  Q^{bc}  \right) \wedge * Q_{bc} - Q_{bc} \wedge \left( \iota_a* Q^{bc}  \right)  , \\
    \overset{(2)}{\tau_a}[Q] =& - 2Q_{ab} \wedge *\left( Q^{bc} \wedge e_c \right) - \left[ \iota_a\left( Q^{dc} \wedge e_c \right) \right] \wedge *\left( Q_{db} \wedge e^b \right) \nonumber \\
    &   \qquad \qquad + \left( Q_{db} \wedge e^b \right) \wedge \left[ \iota_a*\left( Q^{dc} \wedge e_c \right)\right]  , \\
    \overset{(3)}{\tau_a}[Q] =& - 2Q^{bc} \wedge *\left( Q_{ac} \wedge e_b \right) - \left[ \iota_a\left( Q^{dc} \wedge e^b \right) \right] \wedge *\left( Q_{db} \wedge e_c \right) \nonumber \\
    & \qquad \qquad + \left( Q_{db} \wedge e_c \right) \wedge \left[ \iota_a*\left( Q^{dc} \wedge e^b \right)\right]  , \\
   \overset{(4)}{\tau_a}[Q] =& - \left( \iota_a  Q \right) \wedge * Q - Q \wedge \left( \iota_a* Q  \right)  , \\
    \overset{(5)}{\tau_a}[Q] =& - Q \wedge *\left( Q_{ab} \wedge e^b \right) - Q_{ab} \wedge *\left( Q \wedge e^b \right) - \left[ \iota_a\left( Q_{bc} \wedge e^c \right) \right] \wedge *\left( Q \wedge e^b \right) \nonumber \\
    & \qquad \qquad + \left( Q \wedge e^b \right) \wedge \left[ \iota_a*\left( Q_{bc} \wedge e^c \right)\right]  ,
 \end{align}
 \end{subequations}
and 
  \begin{subequations} \label{eq:ang-mom-nonmet-sqred}
  \begin{align}
       \overset{(1)}{\Sigma^b{}_a}[Q] =& 2 *Q^b{}_a  ,\\
      \overset{(2)}{\Sigma^b{}_a}[Q] =&  e^b \wedge *\left( Q_{ac} \wedge e^c \right) +  e_a \wedge *\left( Q^{bc} \wedge e_c \right)  ,\\
      \overset{(3)}{\Sigma^b{}_a}[Q] =&  e^c \wedge *\left( Q_{ac} \wedge e^b \right) +  e_c \wedge *\left( Q^{bc} \wedge e_a \right), \\
    \overset{(4)}{\Sigma^b{}_a}[Q] =& 2 \delta^b_a *Q  ,\\
    \overset{(5)}{\Sigma^b{}_a}[Q] =& \delta^b_a e^c \wedge *\left( Q_{cd} \wedge e^d \right) + \frac{1}{2} \left[  e_a \wedge *\left( Q \wedge e^b \right) +  e^b \wedge *\left( Q \wedge e_a \right) \right]  .
  \end{align}
   \end{subequations}
The field equation governing dynamics of gravity is obtained by eliminating $D\lambda_a$ in (\ref{eq:stpg-field-eqn1}) with help of the equation (\ref{eq:stpg-field-eqn2}) as
 \begin{align}
        \iota_b D \Sigma^b{}_a [Q] + \tau_a[Q]  + \Lambda *e_a = \iota_b D \Sigma^b{}_a[mat] + \tau_a[mat] . \label{eq:stpg-field-eqn}
 \end{align}
While obtaining $D\lambda_a$ we used the results $D(D\rho^b{}_a)=R^b{}_c \wedge \rho^c{}_a - R^c{}_a \wedge \rho^b{}_c=0$ and $De^a=T^a =0$. Then, if need, the Lagrange multipliers can be determined from the equations (\ref{eq:stpg-field-eqn1}) and (\ref{eq:stpg-field-eqn2}).  The configuration below is the GR-equivalent values (for $n \geq 2$)
 \begin{align}
     c_1=\frac{1}{2\kappa} , \qquad c_2= -\frac{1}{\kappa} , \qquad c_3=0, \qquad c_4=-\frac{1}{2\kappa} ,\qquad c_5= \frac{1}{\kappa} .
 \end{align}
We end here with the similar remarks as in the case of metric teleparallelism. 
The number of independent components in the fields is always equal to the number of independent field equations. In symmetric teleparallelism this number is $n(n+3)/2$, which is easily seen in the coordinate frame, wherein the degrees of freedom are packed into the metric and into the flat and torsion-free connection generated by the $n$ diffeomorphisms available in $n$ dimensions. The propagating degrees of freedom have been checked in the generic theory in $n$ dimensions \cite{Conroy:2017yln}. Again, dynamical degrees of freedom may exist in $n=2$ and $n=3$ theories \cite{adak2008,adak-ozdemir:2023}. There are papers discussing the symmetric teleparallel gravity by following the steps defined above, e.g. \cite{adak2013ijmpa,adak2018}. 
 
%We make similar remarks with above. (i) In two dimensions while the number of independent components of nonmetricity is six, that of the field equation is four. That is, unlike general relativity, there are some dynamical degrees of freedom here \cite{adak2008}. (ii) In three dimensions the number of independent components of nonmetricity is 18 and that of the field equation is 9. Thus, unlike general relativity, there are some dynamical degrees of freedom here.  

\subsection{General teleparallel geometry and gravity}

Although we could not solve analytically the connection in terms of metric functions like we do in the Riemannian geometry, we have been able to manage it by using gauge freedom. For practical aim we started in the orthonormal frame and in the coordinate frame for the metric teleparallelism and the symmetric teleparallelism, respectively. Now, we  start in the mixed frame and apply the constraints $Q_{AB} \neq 0$, $ T^A \neq 0$, ${R^A}_B =0$ in the Cartan equations (\ref{eq:cartan-mixedframe}). Again we can not solve ${\omega^A}_B$ analytically because of the third equation. General method is to make ansatz for metric and connection independently, then make them satisfy the constraints above. 

Instead, we can apply to the gauge freedom in the mixed frame by choosing the affine connection zero,
 \begin{align}
   {\omega^A}_B = 0 \qquad \Rightarrow \qquad Q_{AB} = - \frac{1}{2} d g_{AB} \neq 0, \qquad T^A = de^A \neq 0 , \qquad   {R^A}_B =0 . \label{eq:gtpg}
 \end{align}
Then, we are able to obtain the concerned orthonormal quantities via (\ref{eq:orth-to-mix}). For teleparallelism we obtain the inertial connection $\omega^a{}_b = h^a{}_A dh^A{}_b$ where $ h^a{}_A \in GL(n,\mathbb{R})$. The relation between the coframes $ e^a = h^a{}_A e^A$ induces that $h^a{}_A$ is determined by the metric functions, $h^a{}_A = h^a{}_A(g)$. In summary, after determining $e^a(g)$ and $h^a{}_A(g)$ (and $h^A{}_a(g)$), we calculate firstly the full connection, $\omega^a{}_b = h^a{}_A dh^A{}_b$, secondly nonmetricity, $Q_{ab}=\frac{1}{2}(\omega_{ab}+\omega_{ba})$, and torsion, $T^a = de^a + \omega^a{}_b \wedge e^b$, finally substitute all findings into the variational field equations. Consequently, we argue that one can obtain all quantities in the general teleparallel geometry from just metric functions; $g \to e^a(g) \to h^a{}_A(g) \to 
 \omega^a{}_b(g) \to Q_{ab}(g) \; \text{and} \; T^a(g)$.

Here if we choose specially $h^A{}_\alpha(x) = \delta^A_\alpha$ meaning of passage from the mixed frame to the coordinate frame, then the general teleparallel geometry reduces to symmetric teleparallel geometry. On the other hand, if we choose $h^A{}_a(x) = \delta^A_a$ meaning of passage from the mixed frame to the orthonormal frame, it reduces to metric teleparallel geometry.  

The theory of general teleparallel gravity is represented by the total Lagrangian
  \begin{align} \label{eq:tpg-lag}
     L_{GTP} =  L_{T^2} + L_{Q^2} + L_{QT} + \Lambda *1  -  L[mat] +  R^a{}_b \wedge \rho^b{}_a 
 \end{align}
where $L_{T^2}$ is given by (\ref{eq:torsionsquarelagrangian}),  $L_{Q^2}$ is given by (\ref{eq:nonmetricitysquaredlagrangian}) and  Lagrangian of the even parity cross terms is
 \begin{align} \label{eq:crosslagrangian}
     L_{QT} =  \ell_1 (Q^{ab} \wedge e_a \wedge e_c) \wedge * (T^c \wedge e_b) + \ell_2 (Q \wedge e_a) \wedge * T^a 
     + \ell_3 \left(Q_{ab} \wedge e^b\right) \wedge * T^a .
 \end{align}
Here $\ell_1, \ell_2, \ell_3$ are new coupling constants, $\rho^b{}_a$ is Lagrange multiplier $(n-2)$-form constraining the full curvature to zero. Variations with respect to $e^a, \omega^a{}_b, \rho^b{}_a$ generate the field equations of general teleparallel gravity
     \begin{subequations}
  \begin{align}
      \tau_a[Q] + \tau_a[T] + \tau_a[QT] + \Lambda *e_a  &= \tau_a[mat]  , & &\textsc{coframe eqn} \label{eq:tpg-field-eqn} \\
     \Sigma^b{}_a[Q] + \Sigma^b{}_a[T]  + \Sigma^b{}_a[QT] + D\rho^b{}_a &= \Sigma^b{}_a[mat]  , & &\textsc{connection eqn} \label{eq:tpg-field-eqn2} \\
      R^a{}_b &=0 , & &\textsc{zero-curvature eqn}
 \end{align}
  \end{subequations}
where  $\tau_a[mat]$ is matter energy-momentum and $\Sigma^b{}_a[mat]$ is matter angular momentum $(n-1)$-forms derived from $L[mat]$ via the relation $\delta L[mat]= \delta e^a \wedge \tau_a[mat] + \delta \omega^a{}_b \wedge \Sigma^b{}_a[mat]$,
  \begin{align} 
\tau_a[QT] = \sum_{i=1}^3 \ell_i \overset{(i)}{\tau_a}[QT] \qquad \text{and} \qquad \Sigma^b{}_a[QT] = \sum_{i=1}^3 \ell_i \overset{(i)}{\Sigma^b{}_a}[QT]
 \end{align} 
together with
   \begin{subequations} \label{eq:enr-mom-cross}
 \begin{align}
    \overset{(1)}{\tau_a}[QT] =& D\left[ e_b \wedge *\left( Q^{bc} \wedge e_c \wedge e_a\right) \right] + e_c \wedge Q_{ab} \wedge *\left( T^c \wedge e^b\right) - e_c \wedge Q^{bc} \wedge *\left( T_a \wedge e_b\right) \nonumber \\
    & + T^c \wedge *\left( Q_{ab} \wedge e^b \wedge e_c \right) - \left[\iota_a\left( T^c \wedge e_b\right)\right] \wedge *\left( Q^{db} \wedge e_d \wedge e_c \right) \nonumber \\
    & - \left( Q^{db} \wedge e_d \wedge e_c \right) \wedge \left[\iota_a * \left( T^c \wedge e_b\right)\right] ,\\
     \overset{(2)}{\tau_a}[QT] =& D*(Q \wedge e_a) -Q \wedge *T_a -(\iota_aT^b) \wedge *(Q \wedge e_b) + (Q \wedge e_b) \wedge (\iota_a*T^b), \\
     \overset{(3)}{\tau_a}[QT] =& D*(Q_{ab} \wedge e^b) - Q_{ab} \wedge *T^b - (\iota_aT^c) \wedge *(Q_{cb} \wedge e^b) + (Q_{cb} \wedge e^b) \wedge (\iota_a*T^c),
 \end{align}
 \end{subequations}
and 
  \begin{subequations} \label{eq:ang-mom-cross}
  \begin{align}
       \overset{(1)}{\Sigma^b{}_a}[QT] =& e^b \wedge e_d \wedge *\left(Q^{dc} \wedge e_c \wedge e_a \right) + \frac{1}{2} \left[ e_a \wedge e_c \wedge * \left( T^c \wedge e^b\right) + e^b \wedge e_c \wedge * \left( T^c \wedge e_a\right) \right] , \\
    \overset{(2)}{\Sigma^b{}_a}[QT] =&  e^b \wedge *(Q \wedge e_a) + \delta^b_a e_c \wedge *T^c , \\
    \overset{(3)}{\Sigma^b{}_a}[QT] =&  e^b \wedge *(Q_{ac} \wedge e^c) + \frac{1}{2}\left[ e^b \wedge *T_a + e_a \wedge *T^b \right].
  \end{align}
   \end{subequations}
While dynamics of gravity is governed by the equation (\ref{eq:tpg-field-eqn}), the other equation (\ref{eq:tpg-field-eqn2}) is used for determining the Lagrange multiplier. The configuration below is the GR-equivalent values (for $n \geq 3$)
 \begin{align} \label{eq:gr-equiv-gtpg}
       &c_1= \frac{1}{2\kappa}, \qquad c_2=-\frac{1}{\kappa} , \qquad c_3=0 , \qquad c_4= -\frac{1}{2\kappa}, \qquad c_5= \frac{1}{\kappa}, \\
      &k_1=0 , \qquad k_2= \frac{1}{4\kappa}, \qquad k_3= -\frac{1}{2\kappa} , \qquad \ell_1=\frac{1}{\kappa}, \qquad \ell_2= 0, \qquad \ell_3 = 0. \nonumber
 \end{align}
We can make the similar remarks as we did above in the more special cases of metric and symmetric teleparallelisms. In general teleparallelism, there are $n(3n+1)/2$ independent components of the fundamental fields and the same number of independent field equations remain in the field equations after the appropriate manipulations. 

%We make similar remarks with above. (i) In two dimensions since the factors of $k_2$, $k_3$ and $\ell_1$ contain 3-forms, they vanish. While the number of total independent components is 8 (6 from $Q_{ab}$ plus 2 from $T^a$), that of the field equation is 4. That is, unlike in general relativity, there are some dynamical degrees of freedom. (ii) In three dimensions the total numbers of independent components is 27 (18 from $Q_{ab}$ plus 9 from $T^a$), that of the field equation is 9. Therefore, unlike in general relativity, there are some propagating degrees of freedom in free space.

\subsubsection{Example in two dimensions}

Let us give a simple example in two dimensions to make the above ideas more concrete and understandable. We start our algorithm. 

\noindent
{\it Step 1:} Make a static metric ansatz in the coordinate chart $x^\alpha=(t,r)$
 \begin{align}
  ds^2 = -f^2(r) dt^2 + g^2(r) dr^2  
 \end{align}
where $f(r)$ and $g(r)$ are the metric functions.

\noindent
{\it Step 2:} Write the orthonormal covariant components of metric and coframe, $ds^2=\eta_{ab} e^a \otimes e^b$,
  \begin{align}
         \eta_{ab} = \begin{bmatrix}
         -1 & 0 \\
         0 & 1
     \end{bmatrix}  ,  \qquad 
      e^a = \begin{bmatrix}
          f dt \\
          g dr
      \end{bmatrix} .
  \end{align}

\noindent
{\it Step 3:} Write the mixed covariant components of metric and coframe, $ds^2=g_{AB} e^A \otimes e^B$,
 \begin{align}
     g_{AB} = \begin{bmatrix}
         -1 & 0 \\
         0 & g^2
     \end{bmatrix} , \qquad
      e^A = \begin{bmatrix}
          f dt \\
          dr
      \end{bmatrix}.
  \end{align}

\noindent
{\it Step 4:} Determine the zweibein and the inverse via $e^a = h^a{}_A e^A$ and $e^A = h^A{}_a e^a$,
 \begin{align}
      h^a{}_A = \begin{bmatrix}
         1 & 0 \\
         0 & g
     \end{bmatrix} , \qquad
      h^A{}_a = \begin{bmatrix}
         1 & 0 \\
         0 & 1/g
     \end{bmatrix} .
 \end{align}
 
\noindent
{\it Step 5:} Compute the orthonormal affine connection 1-form from $\omega^a{}_b = h^a{}_A dh^A{}_b$,
 \begin{align}
     \omega^a{}_b =
     \begin{bmatrix}
        0 & 0 \\
         0 & -{e^1 g'}/{g^2}
     \end{bmatrix}  .
 \end{align}
After this step we extensively used the computer algebra systems REDUCE and its exterior algebra package EXCALC \cite{hearn-2004,schrufer-2004}.

\noindent
{\it Step 6:} Calculate the orthonormal nonmetricity 1-form from $Q_{ab}= (\omega_{ab}+\omega_{ba})/2$, the orthonormal torsion 2-form from $T^a = de^a + \omega^a{}_b \wedge e^b$ and the orthonormal full curvature 2-form from $R^a{}_b = d\omega^a{}_b + \omega^a{}_c \wedge \omega^c{}_b$,  
 \begin{align}
     Q_{ab} =
     \begin{bmatrix}
        0 & 0 \\
        0 & -({g'}/{g^2})e^1
     \end{bmatrix}  , \qquad 
    T^a = \begin{bmatrix}
         -({f'}/{fg})*1   \\
         0 
     \end{bmatrix}  , \qquad
     R^a{}_b = \begin{bmatrix}
         0 & 0 \\
         0 & 0
     \end{bmatrix}.
 \end{align}
Thus, all the quantities are written in terms of only metric functions. Now we substitute the orthonormal quantities, $e^a, \omega^a{}_b, Q_{ab}, T^a$, into the field equation (\ref{eq:tpg-field-eqn}) in the empty space, i.e. $L[mat]=0$. Thus, we obtain two second order coupled nonlinear differential equations for two unknowns $f(r)$ and $g(r)$. Two classes of solutions are the following.
 
 \bigskip
 
\noindent
{\it Class 1:} Under the constraints $c_1+c_4=0$, $\ell_2=0$, and with the redefinition $\Lambda = k_1 m^2$ we obtained  
  \begin{align}
    f(r) = e^{-m/r} \qquad \text{and} \qquad g(r) = \frac{1}{r^2} .
 \end{align}
By using Taylor expansion $ e^{-m/r} \simeq 1 - {m}/{r}$ we deduce that $m=\sqrt{\Lambda/k_1}$ is mass parameter. There is a singularity at $r=0$, but it does not look an essential singularity because of 
 \begin{align}
      T^a \wedge *T_a = m^2 *1 \qquad \text{and} \qquad  Q_{ab} \wedge *Q^{ab}  = 4r^2*1 .
 \end{align}
We have checked that the covariant exterior derivative of the coframe equation (\ref{eq:tpg-field-eqn}) vanishes for this solution. But at the same time we have to calculate the covariant exterior derivative of the connection equation (\ref{eq:tpg-field-eqn2}) by noticing $D(D\rho^b{}_a)= R^b{}_c \wedge \rho^c{}_a-R^c{}_a \wedge \rho^b{}_c =0$ meaning $D\text{(\ref{eq:tpg-field-eqn2})}=0$. Then we obtain extra constraints on coupling constants. Finally together with dimensional ones $k_2=0$, $k_3=0$, $\ell_1=0$ there are also $c_4=-c_1$, $c_5=2c_1$, $\ell_2=0$, $\ell_3=-2k_1$. The remaining four free coupling constants are $k_1$, $c_1$, $c_2$, $c_3$.

\bigskip
 
\noindent
{\it Class 2:} Under the constraints $c_1+c_4=0$ and $\ell_2 =0$, and with the redefinition $\Lambda = 4k_1m^2n^2$, we found  
  \begin{align}
    f(r) = \left( 1 -\frac{2m}{r} \right)^n \qquad \text{and} \qquad g(r)= \frac{1}{ r^2 \left(1 - \frac{2m}{r} \right)} . 
 \end{align}
By using Taylor expansion $ \left( 1 -\frac{2m}{r} \right)^n \simeq 1 - {2nm}/{r}$ we deduce that $2nm= \sqrt{\Lambda /k_1}$ is mass parameter. There are two singularities at points $r=0$ and $r=2m$, but they seem like coordinate singularities because of 
 \begin{align}
      T^a \wedge *T_a = (2nm)^2 *1 \qquad \text{and} \qquad  Q_{ab} \wedge *Q^{ab}  = 4(r-m)^2*1 .
 \end{align}
Again it is seen that $D$ of (\ref{eq:tpg-field-eqn}) is zero for this solution. But $D\text{(\ref{eq:tpg-field-eqn2})}=0$ yields extra constraints on coupling constants. Finally, as with dimensional constraints $k_2=0$, $k_3=0$, $\ell_1=0$ there are also $c_4=-c_1$, $c_5=2c_1$, $\ell_2=0$, $ \ell_3=-2k_1$. The remaining four free coupling constants are $k_1$, $c_1$, $c_2$, $c_3$.

\subsubsection{Example in three dimensions}

Let us consider circularly symmetric rotating metric in three dimensions. Again we start the algorithm. 

\noindent
{\it Step 1:} Make a metric ansatz in the coordinate chart $x^\alpha=(t,r, \phi)$
 \begin{align}
  ds^2 = -f^2(r) dt^2 + g^2(r) dr^2 + r^2 \left[ w(r) dt + d\phi \right]^2 
 \end{align}
where $f(r)$, $g(r)$ and $w(r)$ are the metric functions.

\noindent
{\it Step 2:} Write the orthonormal covariant components of metric and coframe, $ds^2=\eta_{ab} e^a \otimes e^b$,
  \begin{align}
         \eta_{ab} = \begin{bmatrix}
         -1 & 0 & 0\\
         0 & 1 & 0 \\
         0 & 0 & 1
     \end{bmatrix}  ,  \qquad 
      e^a = \begin{bmatrix}
          f dt \\
          g dr \\
          r \left( w dt + d\phi \right)
      \end{bmatrix} .
  \end{align}

\noindent
{\it Step 3:} Write the mixed covariant components of metric and coframe, $ds^2=g_{AB} e^A \otimes e^B$,
 \begin{align}
     g_{AB} = \begin{bmatrix}
         -1 & 0 &0\\
         0 & g^2 &0 \\
         0 & 0 & 1
     \end{bmatrix} , \qquad
      e^A = \begin{bmatrix}
          f dt \\
          dr \\
          r \left( w dt + d\phi \right)
      \end{bmatrix}.
  \end{align}

\noindent
{\it Step 4:} Determine the dreibein and the inverse via $e^a = h^a{}_A e^A$ and $e^A = h^A{}_a e^a$,
 \begin{align}
      h^a{}_A = \begin{bmatrix}
         1 & 0 &0\\
         0 & g &0\\
         0 & 0 & 1
     \end{bmatrix} , \qquad
      h^A{}_a = \begin{bmatrix}
         1 & 0 & 0\\
         0 & 1/g & 0\\
         0 & 0 & 1
     \end{bmatrix} .
 \end{align}
 
\noindent
{\it Step 5:} Compute the orthonormal affine connection 1-form from $\omega^a{}_b = h^a{}_A dh^A{}_b$,
 \begin{align}
     \omega^a{}_b =
     \begin{bmatrix}
        0 & 0 & 0 \\
         0 & -{e^1 g'}/{g^2} &0 \\
         0 & 0 & 0
     \end{bmatrix}  .
 \end{align}

\noindent
{\it Step 6:} Calculate the orthonormal nonmetricity 1-form from $Q_{ab}= (\omega_{ab}+\omega_{ba})/2$, the orthonormal torsion 2-form from $T^a = de^a + \omega^a{}_b \wedge e^b$ and the orthonormal full curvature 2-form from $R^a{}_b = d\omega^a{}_b + \omega^a{}_c \wedge \omega^c{}_b$,  
 \begin{align}
     Q_{ab} =
     \begin{bmatrix}
        0 & 0 & 0\\
        0 & -\frac{g'e^1}{g^2} &0\\
        0 & 0 & 0
     \end{bmatrix}  , \qquad 
    T^a = \begin{bmatrix}
         -\frac{f' e^{01}}{fg}    \\
         0 \\
         \frac{f e^{12} - r^2 w' e^{01}}{rfg}
     \end{bmatrix}  , \qquad
     R^a{}_b = \begin{bmatrix}
         0 & 0 & 0 \\
         0 & 0 & 0 \\
         0 & 0 & 0
     \end{bmatrix}.
 \end{align}
Please, pay attention again that all quantities are written in terms of only metric functions. Now, we substituted the orthonormal quantities, $e^a, \omega^a{}_b, Q_{ab}, T^a$, into the field equation (\ref{eq:tpg-field-eqn}) in the empty space, i.e. $L[mat]=0$. Thus, we obtained second order coupled differential equations for $f(r)$, $g(r)$ and $w(r)$. Since they are long and complicated we did not write them down. But we could manage finding two classes of solution by the computer algebra systems REDUCE \cite{hearn-2004} and its exterior algebra package EXCALC \cite{schrufer-2004}.

 \bigskip
 
\noindent
{\it Class 1:} Under the constraints $k_1=0$, $c_1+c_4=0$, $\ell_2=0$, we obtained  
  \begin{align}
    f(r) = \frac{1}{g(r)} = \sqrt{m- \frac{\Lambda}{2k_3} r^2} \qquad \text{and} \qquad w(r) = w_0 
 \end{align}
where $m$ and $w_0$ are integration constants. There seems to be one singular point at $r= \sqrt{2mk_3/\Lambda}$. But we find two singular points at $r=0$ and $ r= \sqrt{2mk_3/\Lambda}$ by looking at the invariants  
  \begin{subequations}
 \begin{align}
      T^a \wedge *T_a &= \frac{2m^2 k_3^2 -2 mk_3 \Lambda r^2 + \Lambda^2 r^4}{k_3 r^2 (2m k_3 - \Lambda r^2)} *1 ,\\  
      Q_{ab} \wedge *Q^{ab} &= \frac{\Lambda r^2 }{k_3 (2m k_3 - \Lambda r^2)} *1.
 \end{align}
 \end{subequations}
It is verified that $D$ of (\ref{eq:tpg-field-eqn}) is zero for this solution as well. Furthermore, by using the result $D(D\rho^b{}_a) = R^b{}_c \wedge \rho^c{}_a - R^c{}_a \wedge \rho^b{}_c =0$ we compute the covariant exterior derivative of (\ref{eq:tpg-field-eqn2}) and found extra constraint on the coupling constants. Final situation is below
 \begin{align}
     k_1=0, \quad c_4=-c_1, \quad \ell_2=0, \quad \ell_3=\ell_1 +2c_1 -c_5 + 2k_3,
 \end{align}
where $k_2$, $k_3$, $c_1$, $c_2$, $c_3$, $c_5$, $\ell_1$ are seven free parameters.

 \bigskip
 
\noindent
{\it Class 2:} Under the constraints 
  \begin{align}
      k_1=0, \qquad 2k_2 +k_3=0, \qquad c_1 + c_4=0, \qquad \ell_2=0  \label{eq:const-3d-btz}
  \end{align}
we found
  \begin{align}
    f(r) = \frac{1}{g(r)} = \sqrt{m + \frac{w_1^2}{r^2} - \frac{\Lambda}{2k_3} r^2} \qquad \text{and} \qquad w(r) = \frac{w_1}{r^2} 
 \end{align}
where mass parameter $m$ and rotation parameter $w_1$ are integration constants. This is BTZ metric. We have checked that $D$ of (\ref{eq:tpg-field-eqn}) vanishes for this solution too.  $D\text{(\ref{eq:tpg-field-eqn2})}=0$ reduced the number of free parameters,
 \begin{align}
     k_1=0, \quad k_3=-2k_2, \quad  c_4=-c_1, \quad  c_5=2c_1, \quad \ell_1=4k_2, \quad \ell_2=0, \quad \ell_3=0,
 \end{align}
where $k_2$, $c_1$, $c_2$, $c_3$  are four free parameters. There are three singular points at $r=0$ and $r= \sqrt{\frac{k_3}{\Lambda} \left( m \pm \sqrt{m^2 + \frac{2k_3 w_1^2}{k_3}} \right)}$. They are also singular points of the invariants $T^a \wedge *T_a$ and $Q_{ab} \wedge *Q^{ab}$. We notice that the constraints (\ref{eq:const-3d-btz}) satisfy those (\ref{eq:gr-equiv-gtpg}).

\subsubsection{Example in four dimensions}

Now, we want to give more realistic solutions to our consideration. Firstly we consider the Kerr-de Sitter metric and then Reissner-Nordström metric. 

\noindent
{\it Step 1:} Write the Kerr-de Sitter metric in Boyer-Lindguist coordinates $x^\alpha = (t,r,\theta,\phi)$
 \begin{align}
     ds^2= -\frac{1}{\Omega^2 \Sigma(r,\theta)} \left[ \Delta(r) -j_0^2 \Pi(\theta) \sin^2\theta \right] dt^2 + \frac{2j_0\sin^2\theta}{\Omega^2 \Sigma(r,\theta)} \left[ \Delta(r) -  (r^2 +j_0^2) \Pi(\theta) \right] dt d\phi \nonumber \\
     +\frac{\Sigma(r,\theta)}{\Delta(r)} dr^2 + \frac{\Sigma(r,\theta)}{\Pi(\theta)} d\theta^2 + \frac{\sin^2\theta}{\Omega^2 \Sigma(r,\theta)} \left[  (r^2 +j_0^2)^2 \Pi(\theta) - \Delta(r) j_0^2 \sin^2\theta \right] d\phi^2 
 \end{align}
where $\Delta(r)$, $\Sigma(r,\theta)$, $\Pi(\theta)$ are metric functions, $j_0$ is rotation parameter, $\Omega(\Lambda_0, j_0)$ is a parameter containing cosmological constant and rotation parameter. 

\noindent
{\it Step 2:} Write metric in the orthonormal frame, $g=\eta_{ab} e^a \otimes e^b$, 
  \begin{align}
  \eta_{ab}= \begin{bmatrix}
      -1 & 0 & 0 & 0 \\
      0  & 1 & 0 & 0 \\
       0 & 0 & 1 & 0 \\
       0 & 0 & 0 & 1
  \end{bmatrix} , \qquad 
   e^a =
    \begin{bmatrix}
     \sqrt{\frac{\Delta}{\Sigma \Omega^2}} \left( dt - j_0 \sin^2\theta d\phi \right)  \\
      \sqrt{\frac{\Sigma}{\Delta}} dr  \\
      \sqrt{\frac{\Sigma}{\Pi}} d\theta  \\
     \sqrt{\frac{\Pi}{\Sigma \Omega^2}} \sin\theta \left[  -j_0 dt + (r^2 + j_0^2) d\phi \right]  
     \end{bmatrix}.
 \end{align}

\noindent
{\it Step 3:} Write metric in the mixed frame, $g=g_{AB} e^A \otimes e^B$,
   \begin{align}
  g_{AB}= \begin{bmatrix}
      -\frac{\Delta}{\Sigma \Omega^2} & 0 & 0 & 0 \\
      0  & 1 & 0 & 0 \\
       0 & 0 & 1 & 0 \\
       0 & 0 & 0 & \frac{\Pi \sin^2\theta}{\Sigma \Omega^2}
  \end{bmatrix} , \qquad 
   e^A =
    \begin{bmatrix}
       dt - j_0 \sin^2\theta d\phi  \\
      \sqrt{\frac{\Sigma}{\Delta}} dr  \\
      \sqrt{\frac{\Sigma}{\Pi}} d\theta  \\
        -j_0 dt + (r^2 + j_0^2) d\phi 
     \end{bmatrix}.
 \end{align}

\noindent
{\it Step 4:} Determine the vielbein, $e^a = h^a{}_A e^A$, and the inverse, $e^A = h^A{}_a e^a$,
  \begin{align}
     h^a{}_A =
      \begin{bmatrix}
         \sqrt{\frac{\Delta}{\Sigma \Omega^2}} & 0 & 0 & 0 \\
         0 & 1 & 0 & 0 \\
         0 & 0 & 1 & 0 \\
         0 & 0 & 0 & \sqrt{\frac{\Pi}{\Sigma \Omega^2}} \sin\theta 
      \end{bmatrix} , \qquad 
      h^A{}_a =
      \begin{bmatrix}
         \sqrt{\frac{\Sigma \Omega^2}{\Delta}} & 0 & 0 & 0 \\
         0 & 1 & 0 & 0 \\
         0 & 0 & 1 & 0 \\
         0 & 0 & 0 & \sqrt{\frac{\Sigma \Omega^2}{\Pi}} \frac{1}{\sin\theta} 
      \end{bmatrix} .
  \end{align}

\noindent
{\it Step 5:} Compute the orthonormal affine connection 1-form from $\omega^a{}_b = h^a{}_A dh^A{}_b$,
 \begin{align}
     \omega^a{}_b =
     \begin{bmatrix}
        \sqrt{\frac{\Delta}{\Sigma \Omega^2}} \, d \!\left(\!\sqrt{\frac{\Sigma \Omega^2}{\Delta}} \right) & 0 & 0 & 0 \\
         0 & 0 & 0 & 0 \\
         0 & 0 & 0 & 0 \\
         0 & 0 & 0 & \sqrt{\frac{\Pi}{\Sigma \Omega^2}} \sin\theta \, d \! \left( \! \sqrt{\frac{\Sigma \Omega^2}{\Pi}} \frac{1}{\sin\theta} \right) \\
     \end{bmatrix}  .
 \end{align}

\noindent
{\it Step 6:} Calculate the orthonormal nonmetricity 1-form from $Q_{ab}= (\omega_{ab}+\omega_{ba})/2$, the orthonormal torsion 2-form from $T^a = de^a + \omega^a{}_b \wedge e^b$ and the orthonormal full curvature 2-form from $R^a{}_b = d\omega^a{}_b + \omega^a{}_c \wedge \omega^c{}_b$,  
 \begin{align}
     Q_{ab} \neq 0, \qquad T^a \neq 0 , \qquad R^a{}_b =0.
 \end{align}
We obtain a very complicated second order coupled partial differential equations. With help of Reduce/EXCALC we could find two classes of exact solutions.

 \bigskip
 
\noindent
{\it Class 1:} Under the constraints
 \begin{align} \label{eq:kerr-constraints}
    k_1=0, \quad k_3=-2k_2,  \quad c_3=-c_1-c_2-2k_2, \quad c_5=-c_4 + 2k_2 , \nonumber \\
     \ell_2=\ell_1 -4k_2, \quad \ell_3= -\ell_1 +4k_2, \quad \Lambda = -4k_2 \Lambda_0,
 \end{align}
the field equations are satisfied by the metric functions
  \begin{subequations}
 \begin{align}
     \Delta(r) &= (r^2 + j_0^2)\left( 1- \frac{\Lambda_0}{3}r^2\right) -2m r, \label{eq:Delta-function2}\\
     \Sigma(r,\theta) &= r^2 + j_0^2 \cos^2\theta ,\\
     \Pi(\theta) &= 1+ \frac{\Lambda_0}{3} j_0^2 \cos^2\theta ,\\
     \Omega &= 1 + \frac{\Lambda_0}{3}j_0^2 .
 \end{align}
 \end{subequations}
It is worthy to remark that since $c_1, c_2, c_4, k_2, \ell_1$ are free parameters the constraints (\ref{eq:kerr-constraints}) are more general than the GR-equivalent values of coupling constants in the equation (\ref{eq:gr-equiv-gtpg}). Again we have checked that covariant exterior derivative of the coframe equation (\ref{eq:tpg-field-eqn}) vanishes for this solution too. Similarly the previous results, $D\text{(\ref{eq:tpg-field-eqn2})}=0$ gives rise to new constraints on the coupling constants,
 \begin{align} \label{eq:kerr-constraints2}
    k_1=0, \quad k_3= -2k_2,  \quad c_3=-c_1-c_2-2k_2, \quad c_4=-2k_2 , \quad c_5= 4k_2 , \nonumber \\
     \ell_1=4k_2, \quad \ell_2=0, \quad  \ell_3 =0, \quad \Lambda = -4k_2 \Lambda_0.
 \end{align}
There are still three free coupling constants, $k_2$, $c_1$, $c_2$. We note that rotating spacetimes in metric teleparallelism has been considered in Ref.\cite{Jarv:2019ctf,Bahamonde:2020snl}.

\bigskip
 
\noindent
{\it Class 2:} Under the constraints
 \begin{align}
      k_1=0, \quad k_3=0 , \quad c_3=-c_1 -c_2, \quad c_5 = -c_4 ,  \quad  \ell_2=\ell_1, \quad \ell_3=-\ell_1 , \quad \Lambda=0,
 \end{align}
the field equations are satisfied by the orthonormal coframe
 \begin{align}
     e^0 = \left( 1- \frac{2m}{r} + \frac{q^2}{r^2} \right)^{1/2} dt, \quad e^1 = \left( 1- \frac{2m}{r} + \frac{q^2}{r^2} \right)^{-1/2} dr, \quad e^2 = r d\theta , \quad e^3 = r\sin\theta d\phi ,
 \end{align}
where $m$ and $q$ are integration constants. Here again there are five free parameters, $k_2$, $c_1$, $c_2$, $c_4$,  $\ell_1$. Again, $D$ of (\ref{eq:tpg-field-eqn}) vanishes for this solution too. But by using the result $D(D\rho^b{}_a) = R^b{}_c \wedge \rho^c{}_a - R^c{}_a \wedge \rho^b{}_c =0$ we calculate $D\text{(\ref{eq:tpg-field-eqn2})}=0$ and found extra constraints on the coupling constants. Final situation is below
 \begin{align}
     k_1=0, \quad k_3=0, \quad  c_3=-c_1-c_2, \quad c_4=0, \quad c_5=0,   \nonumber \\
     \ell_1=0, \quad \ell_2=0, \quad \ell_3=0 , \quad \ell_3=0 , \quad \Lambda=0 .
 \end{align}
At the end $k_2$, $c_1$, $c_2$ are three free parameters. It is interesting to see that although we do not add electromagnetic field to our Lagrangian (\ref{eq:tpg-lag}), this metric is the Reissner-Nordström metric of Einstein-Maxwell theory. 
Similar results \cite{Cembranos:2017pcs} have been found in the symmetric teleparallel modified gravity \cite{Bahamonde:2022esv}.
%This result leads us to think about the question, "May general teleparallel gravity be a unification theory of gravity and electromagnetism?" 
 
\section{Conclusions} \label{sec:conclusion}

\begin{figure}
%\begin{center}
\includegraphics[width=1.0\textwidth,keepaspectratio]{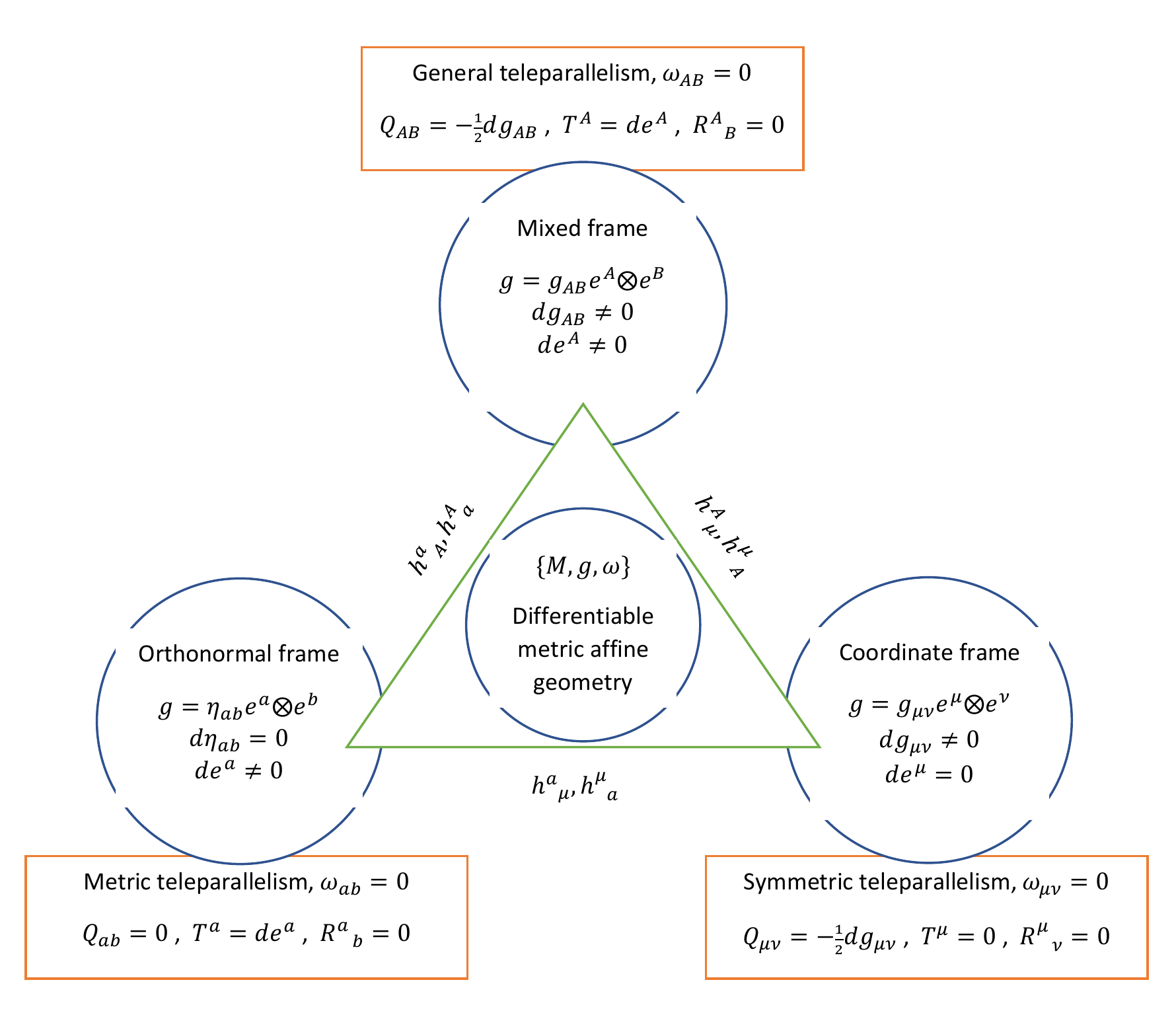} 
\caption{Transformations between the different frames discussed in section \ref{world}, and the descriptions of the three versions of teleparallelism in the respective frames. \label{trinity}}
%\end{center}
\end{figure}

We discussed the teleparallel geometries and theories of gravity developed on them. Three classes of teleparallel geometries were considered. Metric (Weitzenböck) teleparallelism is defined by $Q_{ab} = 0$, $T^a \neq 0$, $R^a{}_b = 0$, symmetric teleparallelism by $Q_{ab} \neq 0$, $T^a = 0$, $R^a{}_b=0$ and general teleparallelism by $Q_{ab} \neq 0$, $T^a \neq 0$, $R^a{}_b=0$. If we match curvature with Riemann, torsion with Cartan, nonmetricity with Weyl, they could also be named as Cartan geometry, Weyl geometry and Cartan-Weyl geometry, respectively. It is well known that the connection is computed analytically from metric in the Riemannian geometry defined by $Q_{ab}=0$, $T^a = 0$, $R^a{}_b \neq 0$. Therefore, one starts with a metric ansatz for gravity models formulated in the Riemannian spacetime.

%Since one can not solve connection analytically in the teleparallel geometries he/she begins by assuming metric and connection independently.
On the other hand, we can also construct teleparallel geometries solely from a metric. There are papers showing that the flat connection could be expressed in terms of metric functions in metric and symmetric teleparallel spacetimes \cite{yakov-itin-1999,adak2013ijmpa,adak-2006-tjp} by virtue of gauge freedom. These works were done in the orthonormal and in the coordinate frames. In this paper, we showed that the connection can be computed from a metric also in the mixed frame, and the gauge freedom allows to exploit the method as well in the general teleparallel spacetime. The relations of the frames and
the different versions of teleparallelism are summarised in the Figure \ref{trinity}.
We wrote down even parity quadratic Lagrangian in each teleparallel spacetime and performed variational calculations explicitly in the language of exterior algebra. Finally, we gave some explicit solutions in two, three and four dimensions to the general teleparallel theory of gravity in order to clarify our arguments on the metric formulation of general teleparallel spacetime. While looking for solutions we used repeatedly the computer algebra system REDUCE \cite{hearn-2004} and its exterior algebra package EXCALC \cite{schrufer-2004}. As seen from two and three dimensional analyses, while there are no dynamical degrees of freedom in general relativity, the Weitzenböck (metric) teleparallel theories of gravity, symmetric teleparallel as well as the general teleparallel theories of gravity can in principle feature propagating modes also in the lower dimensions. %But, we notice that the simplest and richest geometry is symmetric teleparallel geometry.

As a final remark, apart from gravity studies, the mathematical methods and techniques improved here may find opportunity of usage in other fields of physics, such as material physics concerning crystal impurities \cite{roychowdhury-gupta-2017}.

\section*{Acknowledgements}

M.A. stays at Istanbul Technical University (ITU) via a sabbatical leave and thanks the Department of Physics, ITU for warm hospitality. C.P. and M.A. are supported via the project number 2022FEBE032 by the Scientific Research Coordination Unit of Pamukkale University. C.P. thanks TUBITAK (Scientific and Technical Research Council of Turkey) for a grant through TUBITAK 2214-A that makes his stay in the Estonia possible and the Institute of Physics, University of Tartu for warm hospitality. T.S.K. is supported by the Estonian Research Council grant PRG356. T.D. is partially supported by The Turkish Academy of Sciences (TÜBA). We thank to the anonymous referee for useful comments.

\appendix

\section{On tensor formulation}

Let $\{M,g,\nabla\}$ be our spacetime where $M$ is $n$-dimensional differentiable orientable manifold, $g$ is symmetric non-degenerate symmetric second rank covariant metric tensor, $\nabla$ is connection determined by connection 1-forms. We firstly choose the coordinate system $\{\bar{x}^\mu\}$, $\mu=0,1, \cdots , n-1$. Then we can write down metric, nonmetricity 1-form, torsion 2-from, curvature 2-form, respectively,
  \begin{subequations}
      \begin{align}
          g &= \bar{g}_{\mu \nu} d\bar{x}^\mu \otimes d\bar{x}^\nu , \label{eq:metric-bar}\\
          \bar{Q}_{\mu\nu} &= -\frac{1}{2} \left(d \bar{g}_{\mu \nu} - \bar{\omega}^\sigma{}_\mu \bar{g}_{\sigma \nu} - \bar{\omega}^\sigma{}_\nu \bar{g}_{\mu \sigma} \right) , \label{eq:nonmetricity-bar}\\
          \bar{T}^\mu &= d(d\bar{x}^\mu) + \bar{\omega}^\sigma{}_\mu \wedge d\bar{x}^\nu , \label{eq:torsion-bar}\\
          \bar{R}^\mu{}_\nu &= d\bar{\omega}^\mu{}_\nu + \bar{\omega}^\mu{}_\sigma \wedge \bar{\omega}^\sigma{}_\nu \label{eq:curvature-bar},
      \end{align}
  \end{subequations}
where $d$ is the exterior derivative, $\otimes$ is symmetric tensor product, $\wedge$ is exterior product, $d\bar{x}^\mu$ is the coframe (or basis 1-form), $\bar{\omega}^\mu{}_\nu$ is the affine connection 1-form. The last three equations are called the Cartan structure equations.

Now, let ${x}^\mu$ be another coordinate system. We rewrite metric, nonmetricity 1-form, torsion 2-from, curvature 2-form, respectively, in this new coordinate system as
   \begin{subequations}
      \begin{align}
          g &= g_{\mu \nu} dx^\mu \otimes dx^\nu ,\label{eq:metric}\\
          Q_{\mu\nu} &= -\frac{1}{2} \left(d g_{\mu \nu} - \omega^\sigma{}_\mu g_{\sigma \nu} - \omega^\sigma{}_\nu g_{\mu \sigma} \right) , \label{eq:nonmetricity}\\
          T^\mu &= d(dx^\mu) + \omega^\sigma{}_\mu \wedge dx^\nu , \label{eq:torsion}\\
          R^\mu{}_\nu &= d\omega^\mu{}_\nu + \omega^\mu{}_\sigma \wedge \omega^\sigma{}_\nu . \label{eq:curvature}
      \end{align}
  \end{subequations}
Then, let us consider a general coordinate transformation,
  \begin{align}
       x^\mu = x^\mu(\Bar{x}) \qquad \Longleftrightarrow  \qquad \Bar{x}^\mu = \Bar{x}^\mu (x) .
  \end{align}
Under this transformation, the coframe transforms as follows
      \begin{align}
          dx^\mu = \big(\Lambda^{-1}\big)^\mu{}_\nu d\Bar{x}^\nu \qquad \Longleftrightarrow  \qquad d\Bar{x}^\mu = \big(\Lambda\big)^\mu{}_\nu d{x}^\nu 
      \end{align}
where transformation elements satisfying $\big(\Lambda^{-1}\big)^\mu{}_\sigma \big(\Lambda\big)^\sigma{}_\nu = \big(\Lambda\big)^\mu{}_\sigma \big(\Lambda^{-1}\big)^\sigma{}_\nu =\delta^\mu_\nu$ are partial derivatives among two coordinates
  \begin{align}
      \big(\Lambda^{-1}\big)^\mu{}_\nu := \frac{\partial x^\mu}{\partial \Bar{x}^\nu}  \qquad \Longleftrightarrow  \qquad \big(\Lambda\big)^\mu{}_\nu := \frac{\partial \bar{x}^\mu}{\partial x^\nu} .
  \end{align}
These elements are essentially $n\times n$ matrices with entries of zero-forms in the representation furnished by the  general linear group, $\big(\Lambda\big)^\mu{}_\nu \in GL(n,\mathbb{R})$.
Number of the components in generic $GL(n,\mathbb{R})$ matrix is $n^2$, but obviously the general coordinate transformation is determined only by the $n$ independent functions $\bar{x}^\mu(x)$. These generate the integrable subgroup of $GL(n,\mathbb{R})$. 
Besides, independently from the coframe, the affine connection 1-forms transform inhomogeneously 
  \begin{subequations} \label{eq:transformations1}
  \begin{align}
      \omega^\mu{}_\nu &= \big(\Lambda^{-1}\big)^\mu{}_\sigma  \bar{\omega}^\sigma{}_\gamma \big(\Lambda\big)^\gamma{}_\nu + \big(\Lambda^{-1}\big)^\mu{}_\sigma  d\big(\Lambda\big)^\sigma{}_\nu , \label{eq:conn-from-bar} \\
       \bar{\omega}^\mu{}_\nu &= \big(\Lambda\big)^\mu{}_\sigma  \omega^\sigma{}_\gamma \big(\Lambda^{-1}\big)^\gamma{}_\nu + \big(\Lambda\big)^\mu{}_\sigma  d\big(\Lambda^{-1}\big)^\sigma{}_\nu . \label{eq:conn-to-bar}
  \end{align}
  \end{subequations}
Accordingly, metric, nonmetricity, torsion and curvature transform as follows,
 \begin{subequations} \label{eq:transformations2}
     \begin{align}
       g_{\mu \nu} &= \big(\Lambda\big)^\alpha{}_\mu \big(\Lambda\big)^\beta{}_\nu \Bar{g}_{\alpha \beta} & 
 &\Longleftrightarrow  &  \bar{g}_{\mu \nu} &= \big(\Lambda^{-1}\big)^\alpha{}_\mu \big(\Lambda^{-1}\big)^\beta{}_\nu g_{\alpha \beta} \label{eq:transfor-metric}\\
        Q_{\mu \nu} &= \big(\Lambda\big)^\alpha{}_\mu \big(\Lambda\big)^\beta{}_\nu \Bar{Q}_{\alpha \beta} &  &\Longleftrightarrow  &  \bar{Q}_{\mu \nu} &= \big(\Lambda^{-1}\big)^\alpha{}_\mu \big(\Lambda^{-1}\big)^\beta{}_\nu Q_{\alpha \beta} \\  
       T^\mu &= \big(\Lambda^{-1}\big)^\mu{}_\nu \bar{T}^\nu   &  &\Longleftrightarrow  &  \bar{T}^\mu &= \big(\Lambda\big)^\mu{}_\nu T^\nu \\        
        R^\mu{}_\nu &= \big(\Lambda^{-1}\big)^\mu{}_\sigma \bar{R}^\sigma{}_\gamma \big(\Lambda\big)^\gamma{}_\nu  &  &\Longleftrightarrow  &  \bar{R}^\mu{}_\nu &= \big(\Lambda\big)^\mu{}_\sigma R^\sigma{}_\gamma \big(\Lambda^{-1}\big)^\gamma{}_\nu  . 
     \end{align}
 \end{subequations}
Now, in the $x^\mu$ coordinate system, assuming teleparallelism, $R^\mu{}_\nu = 0$, we have
 \begin{align}
     d(dx^\mu) =0 \quad \text{and} \quad  \omega^\mu{}_\nu  = \big(\Lambda^{-1}\big)^\mu{}_\sigma  d\big(\Lambda\big)^\sigma{}_\nu . \label{eq:inertial-connec}
 \end{align}
Furthermore if we write $g_{\mu\nu}(x)$ in terms of a new second rank symmetric covariant tensor $c_{\alpha\beta}(x)$ as
  \begin{align}
      g_{\mu \nu} = \big(\Lambda\big)^\alpha{}_\mu \big(\Lambda\big)^\beta{}_\nu c_{\alpha \beta},
  \end{align}
we can rewrite the Cartan structure equations through (\ref{eq:nonmetricity})-(\ref{eq:curvature}) as the following  
  \begin{align}
      Q_{\mu\nu} = -\frac{1}{2} \big(\Lambda\big)^\alpha{}_\mu \big(\Lambda\big)^\beta{}_\nu dc_{\alpha \beta} , \qquad T^\mu = \big(\Lambda^{-1}\big)^\mu{}_\sigma d\big(\Lambda\big)^\sigma{}_\nu \wedge dx^\nu = 0, \qquad  R^\mu{}_\nu=0 
    \end{align}
Note that the vanishing of torsion is the consequence of $\Lambda^\mu{}_\nu$ being the Jacobian corresponding to a coordinate transformation. A more general element of $GL(n,\mathbb{R})$ would not be a closed $1-form$.   
With help of the rules (\ref{eq:transformations1}) and (\ref{eq:transformations2}) we pass to $\bar{x}^\mu$ coordinate system, and observe that $c_{\mu\nu}(x)= \Bar{g}_{\mu\nu}(\Bar{x}(x))$ and the followings
  \begin{subequations}
      \begin{align}
    \Bar{\omega}^\mu{}_\nu &=0 ,  &    d\Bar{x}^\mu &= \big(\Lambda\big)^\mu{}_\nu  dx^\nu ,  &  \bar{g}_{\mu \nu} &= \big(\Lambda^{-1}\big)^\alpha{}_\mu \big(\Lambda^{-1}\big)^\beta{}_\nu g_{\alpha \beta}, \\
         \bar{Q}_{\mu\nu} &= -\frac{1}{2} d\Bar{g}_{\alpha \beta} , & \bar{T}^\mu &=   0, &  \bar{R}^\mu{}_\nu &= 0. 
      \end{align}
  \end{subequations}
%Since $d(d\Bar{x}^\mu) \neq 0$, we do not call $\Bar{\omega}^\mu{}_\nu =0$ as inertial connection.
Thus, we find that by restricting to symmetric teleparallelism, corresponding precisely to connections generated by an integrable 
$GL(n,\mathbb{R})$ transformation, it is possible formulate {\it any} gravity model solely in terms of the metric components. However, it should be noted that then this metric is not a tensor, we just have its components in a fixed gauge. 

%\bigskip
%\noindent
%Final remark. We can reduce the general teleparallel geometry to the metric teleparallel geometry by choosing $\Bar{g}_{\mu\nu} = \eta_{\mu\nu}$ Minkowski metric and to symmetric teleparallel geometry by choosing $\big(\Lambda\big)^\mu{}_\nu = \delta^\mu_\nu$ Kronecker delta.

\end{document}